\documentclass[prx,reprint,amssymb]{revtex4-2}

\usepackage{xcolor}

\usepackage{bm}

\usepackage{amsmath}  
\usepackage{amsfonts}
\usepackage{graphicx}
\usepackage{hyperref}
\hypersetup{
	colorlinks=true,
	citecolor=blue,
	linkcolor=red,
	urlcolor = blue
}

\begin{document}
	
	\title{Gilbert damping in metallic ferromagnets from Schwinger-Keldysh field theory: Intrinsically nonlocal and nonuniform, and made anisotropic by spin-orbit coupling}
	
	\author{Felipe Reyes-Osorio}
	\author{Branislav K. Nikoli\'c}
	\email{bnikolic@udel.edu}
	\affiliation{Department of Physics and Astronomy, University of Delaware, Newark, DE 19716, USA}
	
	\date{\today}
	
	\begin{abstract}
		Understanding the origin of damping mechanisms in magnetization dynamics of metallic ferromagnets is a fundamental problem for nonequilibrium many-body physics of systems where quantum conduction electrons interact with localized spins assumed to be governed by the classical Landau-Lifshitz-Gilbert (LLG) equation. It is also of critical importance for applications as damping affects energy consumption and speed of spintronic and magnonic devices. Since the 1970s, a variety of  linear-response and scattering theory approaches  have been developed to produce widely used formulas for computation of spatially-independent Gilbert scalar parameter as the magnitude of the Gilbert damping term in the LLG equation. The largely unexploited for this purpose  Schwinger-Keldysh field theory (SKFT) offers additional possibilities, such as to rigorously derive an extended LLG equation by integrating quantum electrons out. Here we derive such equation whose Gilbert damping for metallic ferromagnets is {\em nonlocal}---i.e., dependent on all localized spins at a given time---and {\em nonuniform}, even if all localized spins are collinear and spin-orbit coupling (SOC) is absent. This is in sharp contrast to standard lore, where nonlocal damping is considered to emerge only if localized spins are noncollinear---for such situations, direct comparison on the example of magnetic domain wall shows that SKFT-derived nonlocal damping is an order of magnitude larger than the previously considered one. Switching on SOC makes such nonlocal damping \textit{anisotropic}, in contrast to standard lore where SOC is usually necessary to obtain nonzero Gilbert damping scalar parameter. Our analytical formulas, with their nonlocality being more prominent in low spatial dimensions, are fully corroborated by numerically exact quantum-classical simulations.  
	\end{abstract}
	
	\maketitle
	
	\begin{figure}
		\centering
		\includegraphics[width=\linewidth]{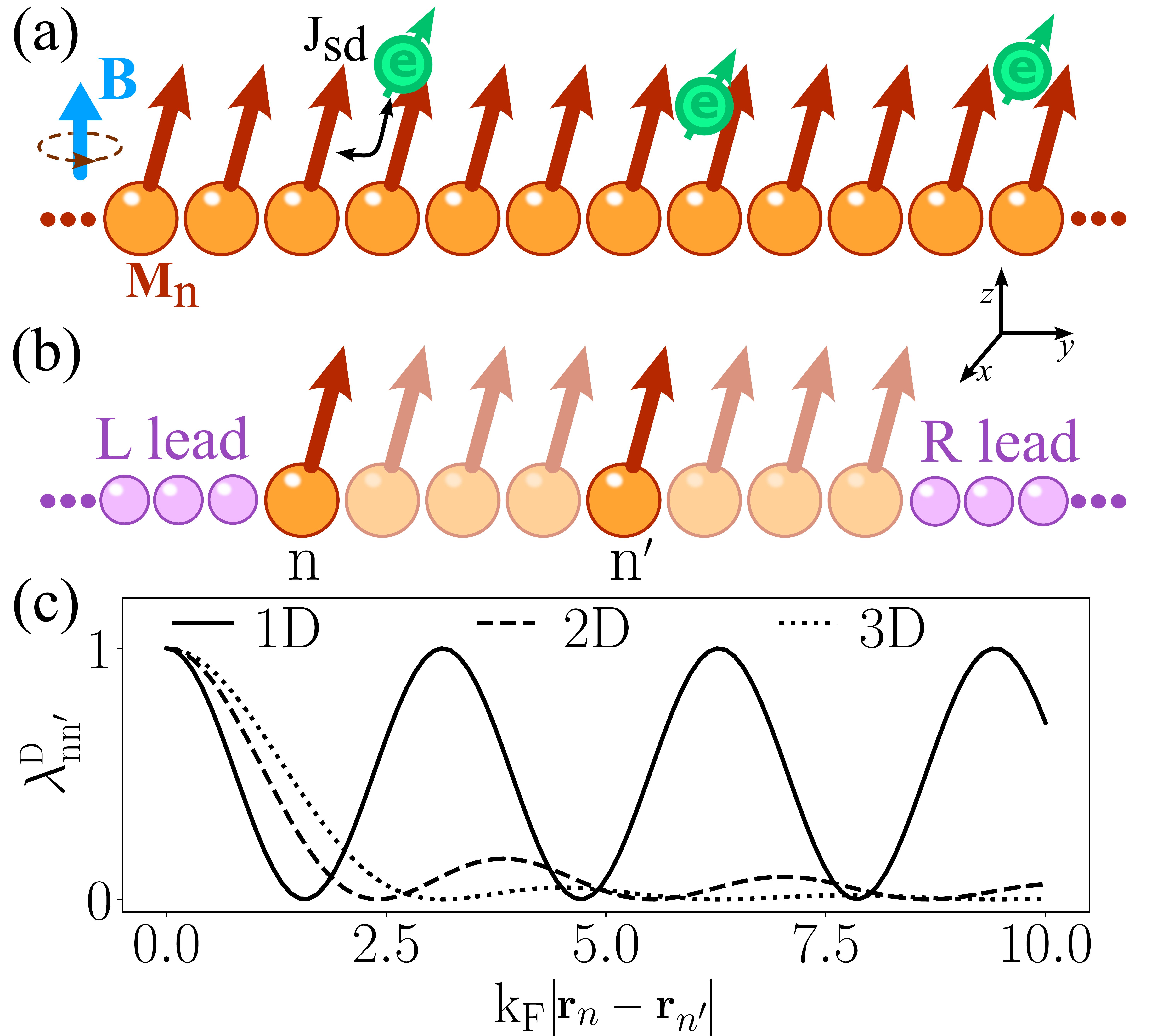}
		\caption{Schematic view of (a) classical localized spins, modeled by unit vectors $\mathbf{M}_n$ (red arrows), within an infinite metallic ferromagnet defined on a cubic lattice in 1D--3D (1D is used in this illustration); or (b) finite-size metallic ferromagnet (central region) attached to semi-infinite NM leads terminating in macroscopic reservoirs, whose difference in electrochemical potentials inject charge current as commonly done in spintronics. The localized spins interact with conduction electron spin  $\langle \hat{\mathbf{s}} \rangle$ (green arrow) via $sd$-exchange of strength $J_{sd}$, while both subsystems can experience external magnetic field $\mathbf{B}$ (blue arrow).  (c) Nonlocal damping $\lambda^D_{nn'}$ [Eq.~\eqref{eq:damping}] obtained from SKFT vs. distance $|\mathbf{r}_n-\mathbf{r}_{n^\prime}|$ between two sites $n$ and $n'$ of the lattice for different dimensionality $D$ of space.}
		\label{fig:fig1}
	\end{figure}
	
	\section{Introduction}
	
	The celebrated Landau-Lifshitz equation~\cite{Landau1935} is the foundation of standard frameworks, such as classical micromagnetics~\cite{Berkov2008,Kim2010} and atomistic spin dynamics~\cite{Evans2014}, for modelling the dynamics of local magnetization within magnetic materials driven by external fields or currents in spintronics~\cite{Berkov2008} and magnonics~\cite{Kim2010}. It considers localized spins as classical vectors $\mathbf{M}(\mathbf{r})$ of fixed length normalized to unity whose rotation around the effective magnetic field $\mathbf{B}_{\rm eff}$ is governed by 
	\begin{equation}\label{eq:oldLLG}
		\partial_t \mathbf{M}=-\mathbf{M} \times \mathbf{B}_{\mathrm{eff}}+\mathbf{M} \times\left(\mathcal{D} \cdot \partial_t \mathbf{M}\right),
	\end{equation}
	where $\partial_t \equiv \partial/\partial t$. Although spin is a genuine quantum degree of freedom, such phenomenological equation can be fully microscopically justified from open quantum many-body system dynamics where $\mathbf{M}(\mathbf{r})$ tracks the trajectories of quantum-mechanical expectation value of localized spin operators~\cite{GarciaGaitan2023} in ferromagnets, as well as in antiferromagnets as long as the spin value is sufficiently large $S>1$. The presence of a dissipative environment in such justification invariably introduces damping mechanisms, which were conjectured phenomenologically in the earliest formulation~\cite{Landau1935}, as well as in later renderings using the so-called Gilbert form of damping~\cite{Saslow2009, Gilbert2004} written as the second term on the right-hand side (RHS) of Eq.~\eqref{eq:oldLLG}. 
	
	The Gilbert damping  $\mathcal{D}$ was originally considered as a spatially uniform scalar $\mathcal{D} \equiv \alpha_G$, or possibly tensor~\cite{Brataas2008,Thonig2014}, dependent on the intrinsic properties of a material. Its typical values are $\alpha_G \sim 0.01$ in standard ferromagnetic metals~\cite{Weindler2014}, or as low as $\alpha_G \sim 10^{-4}$ in carefully designed magnetic insulators~\cite{Soumah2018} and metals~\cite{Schoen2016}. Furthermore, recent extensions~\cite{Zhang2009, Foros2008, Hankiewicz2008, Tserkovnyak2008, Tserkovnyak2009, Kim2012, Yuan2016, Verba2018, Mankovsky2018} of the Landau-Lifshitz-Gilbert (LLG) Eq.~\eqref{eq:oldLLG} for the dynamics of noncollinear magnetization textures find  $\mathcal{D}$ to be a spatially nonuniform and nonlocal tensor
	\begin{equation}\label{eq:dampingTensor}
		\mathcal{D}_{\alpha \beta}=\alpha_G \delta_{\alpha \beta}+\eta \sum_{\beta^\prime}\left(\mathbf{M} \times \partial_{\beta^\prime} \mathbf{M}\right)_\alpha\left(\mathbf{M} \times \partial_{\beta^\prime} \mathbf{M}\right)_\beta,
	\end{equation}
	where $\partial_{\beta^\prime} \equiv \partial/\partial {\beta^\prime}$,  and  $\alpha,\beta,\beta^\prime \in \{x,y,z\}$. 
	
	It is generally believed that $\alpha_G$ is \textit{nonzero only} when SOC~\cite{Mondal2017, Hickey2009} or  magnetic disorder (or both) are present~\cite{Hankiewicz2008,Starikov2010,Starikov2018}. For example, $\alpha_G$ has been extracted from a nonrelativistic expansion of the Dirac equation~\cite{Mondal2017, Hickey2009}, and spin-orbit  coupling (SOC) is virtually always invoked in analytical (conducted for simplistic model Hamiltonians)~\cite{Garate2009a,Garate2009,Ado2020} or first-principles calculations~\cite{Gilmore2007, Ebert2011, Mankovsky2013,Starikov2010,Starikov2018, Hou2019,Guimaraes2019} of $\alpha_G$  via Kubo linear-response~\cite{Kambersky1976,Kambersky1984,Kambersky2007,Ebert2011,Thonig2014} or scattering~\cite{Brataas2008} theory-based formulas.
	
	The second term on the RHS of Eq.~\eqref{eq:dampingTensor} is the particular form~\cite{Zhang2009} of the so-called \textit{nonlocal} (i.e., magnetization-texture-dependent) and \textit{spatially nonuniform} (i.e., position-dependent) damping~\cite{Zhang2009, Baryakhtar1984, Foros2008, Hankiewicz2008, Tserkovnyak2008, Tserkovnyak2009, Kim2012, Yuan2016, Verba2018, Mankovsky2018}. The search for a proper form of nonlocal damping has a long history~\cite{Baryakhtar1984,Yuan2016}. Its importance  has been revealed by experiments~\cite{Weindler2014} extracting very different Gilbert damping  for the same material by using its uniformly precessing localized spins versus dynamics of its magnetic domain walls, as well as in experiments observing wavevector-dependent damping of spin waves~\cite{Li2016}. Its particular form~\cite{Zhang2009} in Eq.~\eqref{eq:dampingTensor} {\em requires} only noncollinear and noncoplanar textures of localized spins, so it can be nonzero even in the absence of SOC, but its presence can greatly enhance its magnitude~\cite{Kim2012} (without SOC, the nonlocal damping in Eq.~\eqref{eq:dampingTensor} is estimated~\cite{Kim2012} to be relevant only for small size \mbox{$\lesssim 1$ nm} noncollinear magnetic textures).
	
	However, recent quantum-classical and numerically exact  simulations~\cite{Sayad2015, Bajpai2019} have revealed that $\alpha_G$ can be nonzero even in the absence of SOC simply because expectation value of conduction electron spin $\langle \hat{\mathbf{s}}\rangle(\mathbf{r})$ is {\em always somewhat behind} $\mathbf{M}(\mathbf{r})$. Such retarded response of electronic spins with respect to motion of classical localized spins, also invoked when postulating extended LLG equation with phenomenological  time-retarded kernel~\cite{Thonig2015}, generates spin torque $\propto \langle \hat{\mathbf{s}}\rangle (\mathbf{r}) \times \mathbf{M}(\mathbf{r})$~\cite{Ralph2008}  and,  thereby, effective Gilbert-like damping~\cite{Thonig2015,Sayad2015,Bajpai2019} that is nonzero in the absence of SOC and operative even if $\mathbf{M}(\mathbf{r})$ at different positions $\mathbf{r}$ are {\em collinear}~\cite{Bajpai2019}. Including SOC in such simulations simply increases~\cite{Suresh2020} the angle between $\langle \hat{\mathbf{s}}\rangle(\mathbf{r})$ and $\mathbf{M}(\mathbf{r})$ and, therefore, the effective damping.
 
    To deepen understanding of the origin of these phenomena observed in numerical simulations, which are analogous to nonadiabatic effects discussed in diverse fields where fast quantum degrees of freedom interact with slow classical ones~\cite{Berry1993, Campisi2012, Thomas2012, Bajpai2020}, requires deriving an analytical expression for Gilbert damping due to interaction between fast conduction electrons and slow localized spins. A rigorous path for such derivation is offered by the Schwinger-Keldysh nonequilibrium field theory (SKFT)~\cite{Kamenev2011} which, however, remains largely unexplored for this problem.  We note that a handful of studies have employed SKFT to study small systems of one or two localized spins~\cite{Onoda2006, Rikitake2005, Fransson2010, Nunez2008, Diaz2012, Leiva2023} as they interact with conduction electrons. While some of these studies~\cite{Onoda2006, Diaz2012, Leiva2023} also arrive at extended LLG equation with nonlocal damping, they are only directly applicable to small magnetic molecules rather than macroscopic ferromagnets in the focus of our study. It is also worth mentioning that an early work~\cite{Rebei2005} did apply SKFT 
    to the same model we are using---electrons whose spins interact via 
    $sd$ exchange interaction with many Heisenberg-exchange-coupled localized spins representing metallic ferromagnet in self-consistent manner---but they did not obtain damping term in their extended Landau-Lifshitz equation, and instead focused on fluctuations in the magnitude of $\mathbf{M}_n$. In contrast, the vectors $\mathbf{M}_n$ are of fixed length in classical micromagnetics~\cite{Berkov2008,Kim2010} and atomistic spin dynamics~\cite{Evans2014}, as well as in our SKFT-derived extended LLG Eq.~\eqref{eq:modifiedLLG} and all other SKFT-based analyses of one or two localized spin problems~\cite{Onoda2006, Rikitake2005, Fransson2010, Nunez2008, Diaz2012, Leiva2023}.
    
   In this study we consider either an infinite [Fig.~\ref{fig:fig1}(a)], or finite [Fig.~\ref{fig:fig1}(b)] but sandwiched between two semi-infinite normal metal (NM) leads terminating in macroscopic electronic reservoirs~\cite{Brataas2008, Nunez2008, Diaz2012}, metallic magnet whose localized spins are coupled by ferromagnetic exchange in equilibrium. The setups in Fig.~\ref{fig:fig1} are of direct relevance to experiments~\cite{Weindler2014,Li2016} on external field [Fig.~\ref{fig:fig1}(a)] or current-driven dynamics~[Fig.~\ref{fig:fig1}(b)] of localized spins in spintronics and magnonics. Our principal result is encapsulated by Fig.~\ref{fig:fig1}(c)---Gilbert damping, due to conduction electron spins not being able to instantaneously follow changes in the orientation of classical localized spins, is always nonlocal and inhomogeneous, with such features becoming more prominent in low-dimensional ferromagnets. This result  is independently confirmed [Fig.~\ref{fig:fig3}] by numerically exact simulations (in one dimension) based on time-dependent nonequilibrium Green's function combined with LLG equation (TDNEGF+LLG) scheme~\cite{Bajpai2019,Petrovic2018,Petrovic2021,Suresh2020}.
	
	We note that conventional linear-response formulas~\cite{Kambersky1976,Kambersky1984,Kambersky2007,Ebert2011,Thonig2014} produce unphysical divergent Gilbert damping~\cite{Guimaraes2019} in a perfectly crystalline magnet at zero temperature. In contrast to previously proposed solutions to this problem---which require~\cite{Costa2015,Edwards2016,Mahfouzi2017a} going beyond the 
	standard picture of electrons that do not interact with each other, while interacting with classical localized spins---our formulas are finite in the clean limit, as well as in the absence of SOC. The scattering theory~\cite{Brataas2008} yields a formula for $\alpha_G$ which is also always finite (in the absence of SOC, it is finite due to spin pumping~\cite{Tserkovnyak2005}). However, that result can only be viewed as a spatial average of our nonlocal damping which cannot produce proper LLG dynamics of local magnetization [Fig.~\ref{fig:pumping}].
	
	The paper is organized as follows. In Sec.~\ref{sec:skft} we formulate the SKFT approach to the dynamics of localized spins interacting with conduction electrons within a metallic ferromagnet. Sections \ref{sec:noSoc} and \ref{sec:soc} show how this approach leads to nonlocal and isotropic, or nonlocal and anisotropic, damping in the presence or absence of SOC, respectively. The SKFT-derived analytical results are corroborated by numerically exact TDNEGF+LLG simulations~\cite{Bajpai2019,Petrovic2018,Petrovic2021,Suresh2020}  in Sec.~\ref{sec:twoSpins}. Then, in Secs.~\ref{sec:pumping} and \ref{sec:dw} we compare SKFT-derived formulas with widely used scattering theory of conventional scalar Gilbert damping~\cite{Brataas2008, Brataas2011, Tserkovnyak2005} or spin-motive force (SMF) theory~\cite{Zhang2009, Yuan2016} of nonlocal damping, respectively. Finally, in Sec.~\ref{sec:dft}, we discuss how to combine our SKFT-derived formulas to first-principles calculations on realistic materials via density functional theory (DFT). We conclude in Sec.~\ref{sec:conclusions}.
	
	\section{Schwinger-Keldysh field theory for metallic ferromagnets}\label{sec:skft}
	
	The starting point of SKFT is the action~\cite{Kamenev2011} of metallic ferromagnet, $S=S_M + S_e$,
	\begin{subequations}
		\begin{eqnarray}
			S_M &=& \int_{\mathcal C}\! dt \sum_n \Big[\partial_t{\mathbf{M}}_n(t)\cdot\mathbf{A}_n - \mathcal{H}[\mathbf{M}_n(t)] \Big], \label{eqs:initialActionM}\\
			S_e &=& \int_{\mathcal C} \! dt \sum_{nn^\prime}\Big[ \bar\psi_n(t) \big(i\partial_t - \gamma_{nn^\prime}\big)\psi_{n^\prime}(t) \label{eqs:initialActionE}\\
			&-& \delta_{nn^\prime} J_{sd}\mathbf{M}_n(t) \cdot \mathbf{s}_{n^\prime}(t)\Big], \nonumber
		\end{eqnarray}
	\end{subequations}
	where $S_M$ is contribution from localized spins and $S_e$ is contribution from conduction electrons. The integration $\int_{\mathcal{C}}$ is along the Keldysh closed contour $\mathcal C$~\cite{Kamenev2011}. Here the subscript $n$ labels the site of a $D$-dimensional cubic lattice; $\partial_t{\mathbf{M}}_n\cdot\mathbf{A}_n$ is the Berry phase term~\cite{Nagaosa2009, Altland2010}; $\mathcal{H}[\mathbf{M}_n]$ is the Hamiltonian of localized spins; $\psi_n = (\psi^\uparrow_n, \psi^\downarrow_n)^T$ is the Grassmann spinor~\cite{Kamenev2011} for an electron at site $n$; \mbox{$\gamma_{nn^\prime}=-\gamma$} is the nearest-neighbor (NN) hopping;  \mbox{$\mathbf{s}_n = \bar\psi_{n}\bm{\sigma}\psi_{n}$} is the electronic spin density, where $\bm{\sigma}$ is the vector of the Pauli matrices; and $J_{sd}$ is the magnitude of $sd$ exchange interaction between flowing spins of conduction electrons and localized spins. For simplicity, we use $\hbar=1$.
	
	The Keldysh contour $\mathcal{C}$, as well as all functions defined on it, can be split into forward ($+$) and backward ($-$) segments~\cite{Kamenev2011}. These functions can, in turn, be rewritten as $\mathbf{M}^\pm_n = \mathbf{M}_{n,c} \pm \frac{1}{2}\mathbf{M}_{n,q}$ for the real-valued localized spins field, and $\psi^\pm_{n} = \frac{1}{\sqrt 2}(\psi_{1,n} \pm \psi_{2,n})$ and $\bar \psi^\pm_n = \frac{1}{\sqrt 2}(\bar \psi_{2,n} \pm \bar \psi_{1,n})$ for the Grassmann-valued fermion fields $\psi_{n}$ and $\bar \psi_{n}$. The subscripts $c$ and $q$ refer to the classical and quantum components of time evolution. This rewriting yields the following expressions for the two actions
	\begin{subequations}\label{eqs:keldyshAction}
		\begin{eqnarray}
			S_M &=& \int\! dt \sum_n M^\alpha_{nq}\left(\epsilon_{\alpha\beta\gamma} \partial_t M^\beta_{n,c} M_{nc}^\gamma + B_{\rm eff}^\alpha[\mathbf{M}_{n,c}]\right), \\
			S_e &=& \int\! dtdt^\prime \sum_{nn^\prime} \bar{\bm \psi}_n^\sigma \big( \check G^{-1}_{nn^\prime}\delta_{\sigma\sigma^\prime} - J_{sd}\check{M}^\alpha_{nn^\prime}\sigma^\alpha_{\sigma\sigma^\prime}  \big) {\bm \psi}_{n^\prime}^{\sigma^\prime}, \phantom{--}
		\end{eqnarray}
	\end{subequations}
	where subscript $\sigma=\uparrow,\downarrow$ is for spin; summation over repeated Greek indices is implied; ${\bm \psi}\equiv(\psi_1, \psi_2)^T$; \mbox{$\mathbf{B}_{\rm eff} = -\delta\mathcal{H}/\delta \mathbf{M}$} is the effective magnetic field; $\epsilon_{\alpha\beta\gamma}$ is the Levi-Civita symbol; and $\check O$ are $2\times 2$ matrices in the Keldysh space, such as
	\begin{equation}
		\check G_{nn^\prime} = \begin{pmatrix}
			G^R & G^K \\
			0     & G^A
		\end{pmatrix}_{nn^\prime}, \quad 
		\check{M}_{nn^\prime}^\alpha = \begin{pmatrix}
			M_{c}  & \frac{M_q}{2} \\
			\frac{M_q}{2}    &  M_c 
		\end{pmatrix}_n^\alpha\delta_{nn^\prime}.
	\end{equation}
	Here $G^{R/A/K}_{nn^\prime}(t, t^\prime)$ are electronic retarded/advanced/Keldysh Green's functions (GFs)~\cite{Kamenev2011} in the real-space representation of sites $n$.
	
	The electrons can be integrated out~\cite{Onoda2006} up to the second order in $J_{sd}$ coupling, thereby yielding an effective action for localized spins only
	\begin{eqnarray}\label{eq:effectiveAction}
		S_{M}^{\rm eff} &=& \int\! dt \sum_n M^\alpha_{n,q}\Big[\epsilon_{\alpha\beta\gamma} \partial_t M^\beta_{n,c} M_{n,c}^\gamma + B_{\rm eff}^\alpha[\mathbf{M}_{n,c}] \nonumber\\
		&+&\int\! dt^\prime \sum_{n^\prime} M^\alpha_{n^\prime,c}(t^\prime) \eta_{nn^\prime}(t, t^\prime)\Big],
	\end{eqnarray}
	where 
	\begin{eqnarray}\label{eq:dissipationKernels}
		\eta_{nn^\prime}(t,t') &  = & iJ_{sd}^2\Big( G^R_{nn^\prime}(t,t^\prime) G^K_{nn^\prime}(t^\prime, t) \nonumber \\
		&& + G^K_{nn^\prime}(t,t^\prime)G^A_{nn^\prime}(t^\prime,t)\Big),
	\end{eqnarray}
	is the non-Markovian time-retarded kernel. Note that terms that are second order in the quantum fluctuations $\mathbf{M}_{n,q}$ are neglected~\cite{Kamenev2011} in order to write Eq.~\eqref{eq:effectiveAction}. The magnetization damping can be explicitly extracted by analyzing the kernel, as demonstrated for different ferromagnetic setups in Secs.~\ref{sec:noSoc} and \ref{sec:soc}.

	\section{Results and discussion}\label{sec:results}
	
	\subsection{Nonlocality of Gilbert damping in metallic ferromagnets in the absence of SOC}\label{sec:noSoc}
	
	Since $\eta_{nn^\prime}(t-t^\prime)$ depends only on the difference $t-t^\prime$, it can be Fourier transformed to energy $\varepsilon$. Thus, the kernel can be written down  explicitly  for low energies as
	\begin{equation}\label{eq:polarizationEnergy}
		\eta_{nn^\prime}(\varepsilon) = J_{sd}^2\frac{i\varepsilon}{2\pi} \sum_{\mathbf{k,q}} e^{i\mathbf{k}\cdot(\mathbf{r}_n - \mathbf{r}_{n^\prime})} e^{i\mathbf{q}\cdot(\mathbf{r}_n - \mathbf{r}_{n^\prime})} A_{\mathbf{k}}(\mu)A_{\mathbf{q}}(\mu),
	\end{equation}
	where $A_{\mathbf{k}}(\mu) \equiv i[G^R_\mathbf{k}(\mu)-G^A_\mathbf{k}(\mu)]$ is the spectral function~\cite{Nunez2008} evaluated at chemical potential $\mu$; $\mathbf{k}$ is a wavevector; and $\mathbf{r}_n$ and $\mathbf{r}_{n^\prime}$ are the position vectors of sites $n$ and $n^\prime$. Equation~\eqref{eq:polarizationEnergy} remains finite in the clean limit and for low temperatures, so it evades unphysical divergences in the linear-response approaches~\cite{Costa2015, Edwards2016, Mahfouzi2017a}. By transforming it back into the time domain, we minimize the effective action in Eq.~\eqref{eq:effectiveAction} with respect to the quantum fluctuations to obtain semiclassical equations of motion for classical localized spins. This procedure is equivalent to the so-called large spin approximation~\cite{Shnirman2015, Verstraten2023} or a one loop truncation of the effective action. The higher order terms neglected in Eq.~\eqref{eq:effectiveAction} contribute a stochastic noise that vanishes in the low temperature and large spin limit. Although the fluctuating effect of this noise can modify the exact dynamics~\cite{Shnirman2015, Leiva2023}, the deterministic regime suffices for a qualitative understanding and is often the main focus of interest~\cite{Hurst2020, Verstraten2023}. 
	
	Thus, we arrive at the following extended LLG equation
	\begin{equation}\label{eq:modifiedLLG}
		\partial_t \mathbf{M}_{n} = -\mathbf{M}_{n} \times \mathbf{B}_{{\rm eff},n}  + \mathbf{M}_n \times \sum_{n^\prime} \lambda^D_{nn^\prime}\partial_t \mathbf{M}_{n^\prime},
	\end{equation}
	where the conventional $\alpha_G \mathbf{M}_{n} \times \partial_t \mathbf{M}_{n}$ Gilbert term is replaced by the second term on the RHS exhibiting nonlocal damping  $\lambda^D_{nn^\prime}$ instead of Gilbert damping scalar parameter $\alpha_G$. A closed expression for $\lambda^D_{nn^\prime}$ can be obtained for one-dimensional (1D), two-dimensional (2D) and three-dimensional (3D) metallic ferromagnets by considering quadratic energy-momentum dispersion of their conduction electrons
	\begin{equation}\label{eq:damping}
		\lambda^D_{nn^\prime} = \begin{cases}
			\frac{2J_{sd}^2}{\pi v_F^2}\cos^2(k_F|\mathbf{r}_n-\mathbf{r}_{n^\prime}|) & {\rm 1D}, \\
			\frac{k_F^2 J_{sd}^2}{2\pi v_F^2}J_0^2(k_F|\mathbf{r}_n-\mathbf{r}_{n^\prime}|) & {\rm 2D}, \\
			\frac{k_F^2 J_{sd}^2}{2\pi v_F^2}\frac{\sin^2(k_F |\mathbf{r}_n-\mathbf{r}_{n^\prime}|)}{|\mathbf{r}_n-\mathbf{r}_{n^\prime}|^2} & {\rm 3D}.
		\end{cases}
	\end{equation}
	Here $k_F$ is the Fermi wavevector of electrons, $v_F$ is their Fermi velocity, and $J_0(x)$ is the 0-th Bessel function of the first kind.
	
	\subsection{Nonlocality and anisotropy of Gilbert damping in metallic ferromagnets in the presence of SOC}\label{sec:soc}
	
	\begin{figure}
		\centering
		\includegraphics[width = \linewidth]{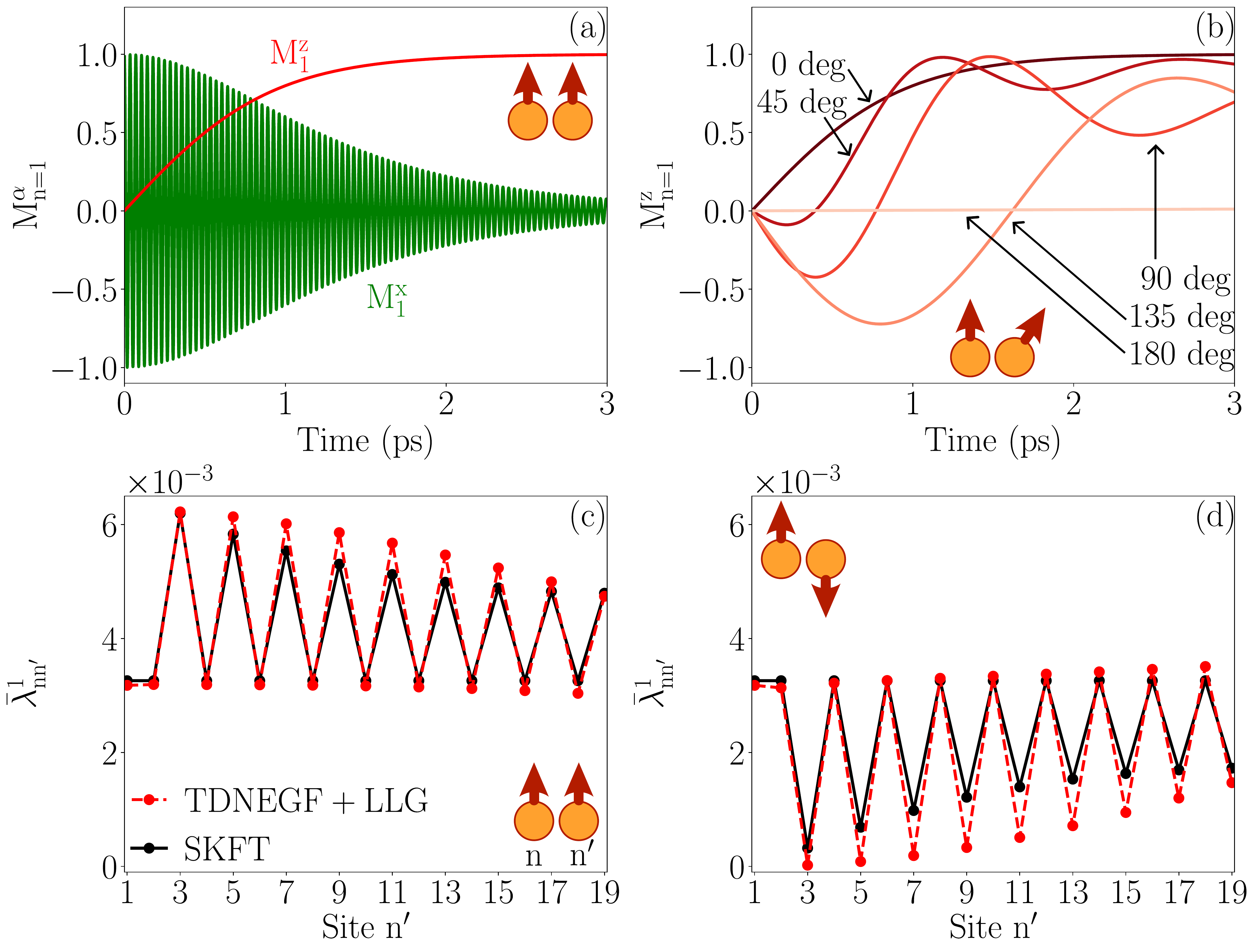}
		\caption{(a) Time evolution of two localized spins $\mathbf{M}_n$, located at sites $n=1$ and $n^\prime=3$ within a chain of 19 sites in the setup of Fig.~\ref{fig:fig1}(b), computed numerically by TDNEGF+LLG  scheme~\cite{Bajpai2019,Petrovic2018,Petrovic2021,Suresh2020}. The two spins are collinear at $t=0$ and point along the $x$-axis, while magnetic field is applied along the $z$-axis. (b) The same information as in panel (a), but for two noncollinear spins with angle $\in \{ 0, 45, 90, 135, 180 \}$ between them. (c) and (d) Effective damping extracted from TDNEGF+LLG simulations (red dashed line) vs. the one from SKFT [black solid line plots 1D case in Eq.~\eqref{eq:damping}] as a function of the site $n^\prime$ of the second spin. The two spins are initially parallel in (c), or antiparallel in (d). The Fermi wavevector of conduction electrons is chosen as $k_F=\pi/2a$, where $a$ is the lattice spacing.}
		\label{fig:fig3}
	\end{figure}

	Taking into account that previous analytical calculations~\cite{Garate2009, Garate2009a, Ado2020} of conventional Gilbert damping scalar parameter always include SOC, often of the Rashba type~\cite{Manchon2015}, in this section we show how to generalize Eq.~\eqref{eq:polarizationEnergy} and nonlocal damping extracted in the presence of SOC. For this purpose, we employ the Rashba Hamiltonian in 1D, with its  diagonal representation given by, $\hat H = \sum_{k\sigma} \varepsilon_{k\sigma} \hat c^\dagger_{k\sigma}\hat c_{k\sigma}$,
	where $\hat c_{k\sigma}^\dagger / \hat c_{k\sigma}$ creates/annihilates an electron with wavenumber $k$ and spin $\sigma$ oriented along the $y$-axis, \mbox{$\varepsilon_{k\sigma} =  -2\gamma\cos k + 2\sigma\gamma_{\rm SO} \sin k$} is the Rashba spin-split energy-momentum dispersion, and $\gamma_{\rm SO}$ is the strength of the Rashba SOC coupling. By switching from second-quantized operators $\hat c_{k\sigma}^\dagger / \hat c_{k\sigma}$ to Grassmann-valued two-component fields~\cite{Altland2010} $\bar{\mathbf{ c}}^\sigma_n/\mathbf{c}^\sigma_n$, where $\mathbf{c}^\sigma_n = (c^\sigma_{1,n}, c^\sigma_{2,n})^T$, we obtain for the electronic action
	\begin{equation}\label{eq:socAction}
		S_e = \int\! dtdt^\prime \sum_{nn^\prime} \bar{\bm c}_n^\sigma \big[ (\check G^{\sigma}_{nn^\prime})^{-1}\delta_{\sigma\sigma^\prime} - J_{sd}\check{M}^\alpha_{nn^\prime}\sigma^{\beta}_{\sigma\sigma^\prime}  \big] {\bm c}_{n^\prime}^{\sigma^\prime}.
	\end{equation}
	Here $\check G^\sigma_{nn^\prime}$ is diagonal, but it depends on spin through $\varepsilon_{k\sigma}$. In addition, $\check M^{x,y,z}_{nn^\prime}$, as the matrix which couples to the same $\sigma^{x,y,z}$ Pauli matrix in electronic action without SOC [Eq.~\eqref{eqs:initialActionE}], is coupled in Eq.~\eqref{eq:socAction} to a different Pauli matrix $\sigma^{y,z,x}$. 
	
	By integrating electrons out up to the second order in $J_{sd}$, and by repeating steps analogous to those of Sec.~\ref{sec:skft} while carefully differentiating the spin-split bands, we find that nonlocal damping becomes anisotropic
	\begin{equation}\label{eq:matrixDamping}
		\lambda_{nn^\prime}^{\rm 1D} = \begin{pmatrix}
			\alpha^\perp_{nn^\prime} & 0 & 0 \\
			0 & \alpha^\parallel_{nn^\prime} & 0 \\
			0 & 0 & \alpha^\perp_{nn^\prime}.
		\end{pmatrix}.
	\end{equation}
	where
	\begin{subequations}\label{eq:dampingSOC}
		\begin{align}
			\alpha^\perp_{nn^\prime} &= \frac{J_{sd}^2}{\pi}\bigg(\frac{\cos^2(k_F^\uparrow |\mathbf{r}_n-\mathbf{r}_{n^\prime}|)}{{v_F^\uparrow}^2} + \frac{\cos^2(k_F^\downarrow |\mathbf{r}_n-\mathbf{r}_{n^\prime}|)}{{v_F^\downarrow}^2} \bigg),\\
			\alpha^\parallel_{nn^\prime} &= \frac{J_{sd}^2}{\pi|v_F^\uparrow v_F^\downarrow|}\Big(\cos\big[(k_F^\uparrow + k_F^\downarrow)|\mathbf{r}_n-\mathbf{r}_{n^\prime}|\big] \label{eq:dampingSOC2}\\
			&+ \cos\big[(k_F^\uparrow - k_F^\downarrow) |\mathbf{r}_n-\mathbf{r}_{n^\prime}|\big]\Big), \nonumber 
		\end{align}
	\end{subequations}
	and $k_F^{\uparrow/\downarrow}$ and $v_F^{\uparrow/\downarrow}$ are the Fermi wavevectors and velocities, respectively, of the Rashba spin-split bands. This means that the damping term in Eq.~\eqref{eq:modifiedLLG} is now given by $\mathbf{M}_n\times\sum_{n^\prime} \lambda_{nn^\prime}^{\rm 1D}\cdot \partial_t\mathbf{M}_{n^\prime}$. 
	
	We note that previous experimental~\cite{Baker2016}, numerical~\cite{Faehnle2008, Thonig2014}, and analytical~\cite{Ado2020, Garate2009, Garate2009a} studies have also found SOC-induced anisotropy of Gilbert damping scalar parameter. However,  our results [Eqs.~\eqref{eq:matrixDamping} and \eqref{eq:dampingSOC}] exhibit additional  feature of nonlocality (i.e., damping at site $n$ depends on spin at site $n^\prime$) and nonuniformity (i.e., dependence on $|\mathbf{r}_n-\mathbf{r}_{n^\prime}|$). As expected from Sec.~\ref{sec:noSoc}, nonlocality persists for $\gamma_{\rm SO}=0$, i.e., $k_F^\uparrow = k_F^\downarrow=k_F$, with $\lambda^{\rm 1D}_{nn^\prime}$ properly reducing to contain $\alpha_{nn^\prime}$ three diagonal elements. Additionally, the damping component $\alpha^\parallel_{nn^\prime}$ given by Eq.~\eqref{eq:dampingSOC2} can take negative values, revealing the driving capability of the conduction electrons (see Sec.~\ref{sec:twoSpins}). However, for realistic small values of $\gamma_{\rm SO}$, the driving contribution of nearby localized spins is likewise small. Furthermore, the decay of nonlocal damping with increasing distance observed in 2D and 3D, together with the presence of intrinsic local damping from other sources, ensures that the system tends towards equilibrium.
	
	\subsection{Comparison of SKFT-derived formulas with numerically exact TDNEGF+LLG simulations}\label{sec:twoSpins}
	
	An analytical solution to Eq.~\eqref{eq:modifiedLLG} can be obtained in few special cases, such as for two exchange-uncoupled localized spins at sites $n=1$ and $n^\prime \neq 1$ within 1D wire placed in an external magnetic field $\mathbf{B}_{\rm ext} = B_{\rm ext} \mathbf{e}_{z}$, on the proviso that the two spins are collinear at $t=0$. The same system can be simulated by TDNEGF+LLG scheme, so that comparing analytical to such numerically exact solution for trajectories $\mathbf{M}_n(t)$ makes it possible to investigate accuracy of our derivation and approximations involved in it, such as: truncation to $J_{sd}^2$ order; keeping quantum fluctuations $\mathbf{M}_{n,q}$ to first order; and low-energy approximation used in Eq.~\eqref{eq:polarizationEnergy}. 
	While such a toy model is employed to verify the SKFT-based derivation, we note that two uncoupled localized spins can also be interpreted as macrospins of two distant ferromagnetic layers within a spin valve for which oscillatory Gilbert damping as a function of distance between the layers was observed experimentally~\cite{Montoya2014}.
	Note that semi-infinite NM leads from the setup in Fig.~\ref{fig:fig1}(b), always used in TDNEGF+LLG simulations to ensure continuous energy spectrum of the whole system~\cite{Petrovic2018, Bajpai2019}, can also be included in SKFT-based derivation by using self-energy $\Sigma_k^{R/A}(\varepsilon)$~\cite{Nunez2008, Ryndyk2016} which modifies the GFs of the central magnetic region in Fig.~\ref{fig:fig1}(b), \mbox{$G^{R/A}_k = (\varepsilon - \varepsilon_k - \Sigma^{R/A}_k)^{-1}$}, where $\varepsilon_k = -2\gamma \cos k$.

	The TDNEGF+LLG-computed trajectory $\mathbf{M}_1(t)$ of localized spin at site $n=1$ is shown in Figs.~\ref{fig:fig3}(a) and \ref{fig:fig3}(b) using two localized spins which are initially collinear or noncollinear, respectively. For the initially parallel [Fig.~\ref{fig:fig3}(a)] or antiparallel localized spins, we can extract Gilbert damping from such trajectories because $M^z_1(t)~=~\tanh\big(\bar \lambda^{\rm 1D}_{nn^\prime}B_{\rm ext}t/(1+ (\bar \lambda^{\rm 1D} _{nn^\prime})^2)\big)$~\cite{Evans2014, Bajpai2019}, where the effective damping is given by $\bar \lambda^{\rm 1D}_{nn^\prime} = \lambda^{\rm 1D}_{00} \pm \lambda^{\rm 1D}_{nn^\prime}$ ($+$ for parallel and $-$ for antiparallel initial condition). The nonlocality of such effective damping in Figs.~\ref{fig:fig3}(c) and \ref{fig:fig3}(d) manifests as its oscillation with increasing separation of the two localized spins. The same result is predicted by the SKFT-derived formula [1D case in Eq.~\eqref{eq:damping}], which remarkably closely traces the numerically extracted $\bar\lambda_{nn^\prime}^{\rm 1D}$ despite approximations involved in SKFT-based analytical derivation. Note also that the two localized spins remain collinear at all times $t$, but damping remains nonlocal. The feature missed by the SKFT-based formula is the decay of $\bar \lambda_{nn^\prime}^{\rm 1D}$ with increasing $|\mathbf{r}_n-\mathbf{r}_{n^\prime}|$, which is present in numerically-extracted effective damping in Figs.~\ref{fig:fig3}(c) and \ref{fig:fig3}(d). Note that effective drastically reduced for antiparallel initial conditions, due to the driving capabilities of the conduction electrons, in addition to their dissipative nature. For noncollinear initial conditions, TDNEGF+LLG-computed trajectories become more complicated [Fig.~\ref{fig:fig3}(b)], so that we cannot extract the effective damping $\lambda^{\rm 1D}_{nn^\prime}$ akin to Figs.~\ref{fig:fig3}(c) and \ref{fig:fig3}(d) for the collinear initial conditions. 
	
	\subsection{Comparison of SKFT-derived formulas with the scattering theory~\cite{Brataas2008} of uniform local Gilbert damping}\label{sec:pumping} 
	
	The scattering theory of Gilbert damping $\alpha_G$ was formulated by studying a single domain ferromagnet in contact with a thermal bath~\cite{Brataas2008}. In such a setup, energy~\cite{Brataas2008} and spin~\cite{Tserkovnyak2005} pumped out of the system by time-dependent magnetization contain information about spin-relaxation-induced bulk~\cite{Brataas2008,Brataas2011} and interfacial~\cite{Tserkovnyak2005} separable contributions to $\alpha_G$,  expressible in terms of the scattering matrix of a ferromagnetic layer attached to two semi-infinite NM leads. For collinear localized spins of the ferromagnet, precessing together as a macrospin,  scattering theory-derived $\alpha_G$ is a spatially-uniform  scalar which can be anisotropic~\cite{Brataas2011}. Its expression is equivalent~\cite{Brataas2011} to Kubo-type formulas~\cite{Kambersky1976,Kambersky1984,Kambersky2007, Thonig2014} in the linear response limit, while offering an efficient algorithm for numerical first-principles calculations~\cite{Starikov2010,Starikov2018} that can include disorder and SOC on an equal footing. 
	
	\begin{figure}
		\centering
		\includegraphics[width = \linewidth]{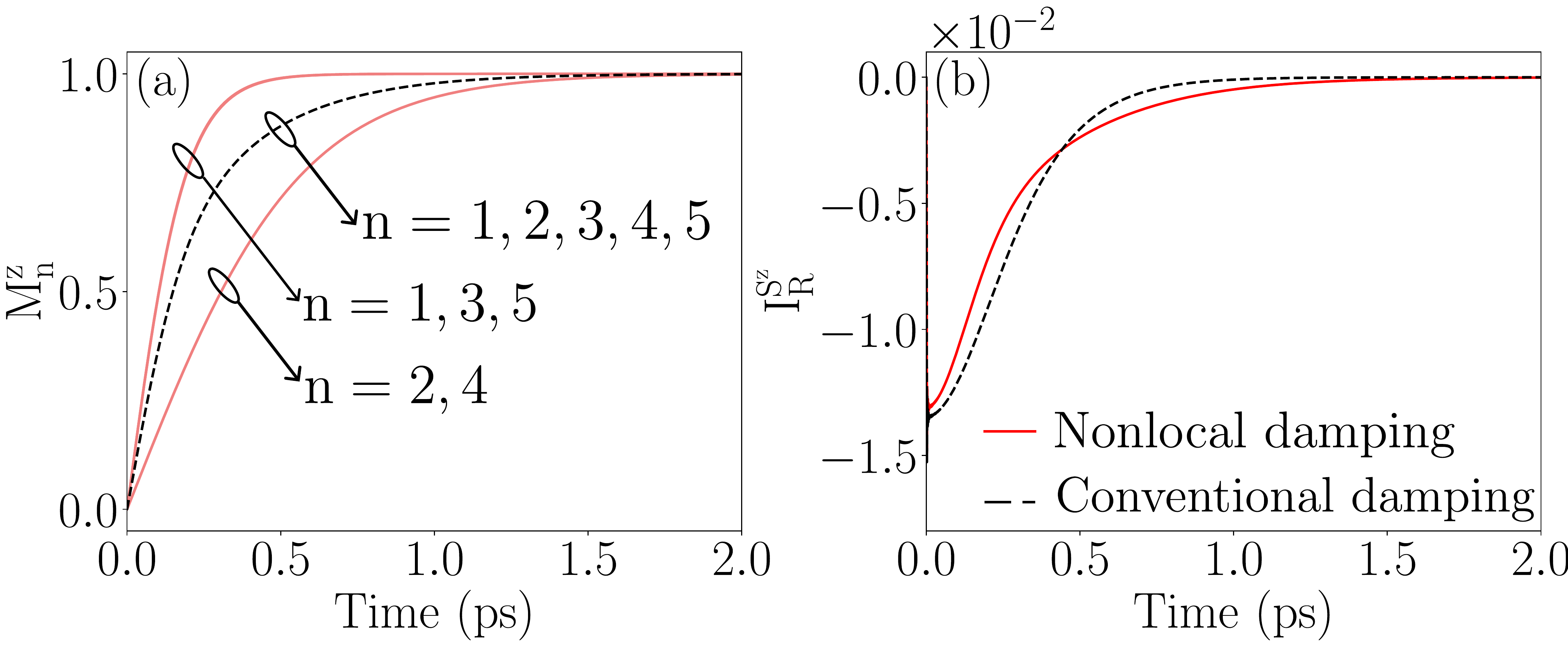}
		\caption{(a) Comparison of trajectories of localized spins $M^z_n(t)$, in the setup of Fig.~\ref{fig:fig1}(b) whose central region is 1D metallic ferromagnet composed of $5$ sites, using LLG Eq.~\eqref{eq:modifiedLLG} with SKFT-derived nonlocal damping (solid red lines) vs. LLG equation with conventional spatially-independent $\alpha_G=0.016$ (black dashed line). This value of $\alpha_G$ is obtained by averaging nonlocal damping over the whole ferromagnet. The dynamics of $\mathbf{M}_n(t)$ is initiated by an external magnetic field along the $z$-axis, while all five localized spins point along the $x$-axis at $t=0$. (b) Comparison of spin current $I^{S^z}_R(t)$ pumped~\cite{Tserkovnyak2005, Petrovic2018, Petrovic2021} by the dynamics of $\mathbf{M}_n(t)$ for the two cases [i.e., nonuniform $\mathbf{M}_n(t)$ for nonlocal vs. uniform $\mathbf{M}_n(t)$ for conventional damping] from panel (a). The Fermi wavevector of conduction electrons is chosen as $k_F=\pi/2a$.}
		\label{fig:pumping}
	\end{figure}

	On the other hand, even if all localized spins are initially collinear, SKFT-derived extended LLG Eq.~\eqref{eq:modifiedLLG} predicts that due to nonlocal damping each localized spin will acquire a \textit{distinct} $\mathbf{M}_n(t)$ trajectory, as demonstrated  by solid red lines in Fig.~\ref{fig:pumping}(a). By feeding these trajectories, which are affected by nonlocal damping [1D case in Eq.~\eqref{eq:damping}] into TDNEGF+LLG simulations, we can compute spin current $I_R^{S^z}(t)$ pumped~\cite{Petrovic2018, Petrovic2021} into the right semi-infinite lead of the setup in Fig.~\ref{fig:fig1}(b) by the dynamics of $\mathbf{M}_n(t)$. A very similar result for pumped spin current is obtained [Fig.~\ref{fig:pumping}(b)] if we feed identical $\mathbf{M}_n(t)$ trajectories [black dashed line in Fig.~\ref{fig:pumping}(a)] from conventional LLG equation with Gilbert damping scalar parameter, $\alpha_G$, whose value is obtained by averaging the SKFT-derived nonlocal damping over the whole ferromagnet. This means that scattering theory of Gilbert damping~\cite{Brataas2008}, which in this example is purely due to interfacial spin pumping~\cite{Tserkovnyak2005} because of lack of SOC and disorder (i.e., absence of spin relaxation in the bulk), would predict a constant $\alpha_G$ that can only be viewed as the spatial average of SKFT-derived nonlocal and nonuniform $\lambda^{\rm 1D}_{nn^\prime}$. In other words, Fig.~\ref{fig:pumping} reveals that different types of microscopic magnetization dynamics $\mathbf{M}_n(t)$ can yield the same total spin angular momentum loss into the external circuit, which is, therefore, insufficient on its own to decipher details (i.e., the proper form of extended LLG equation) of microscopic dynamics of local magnetization.
	
	\subsection{Comparison of SKFT-derived formulas with spin motive force theory~\cite{Zhang2009} and~\cite{Yuan2016} of nonlocal damping}\label{sec:dw}
	
	\begin{figure}
		\centering
		\includegraphics[width = \linewidth]{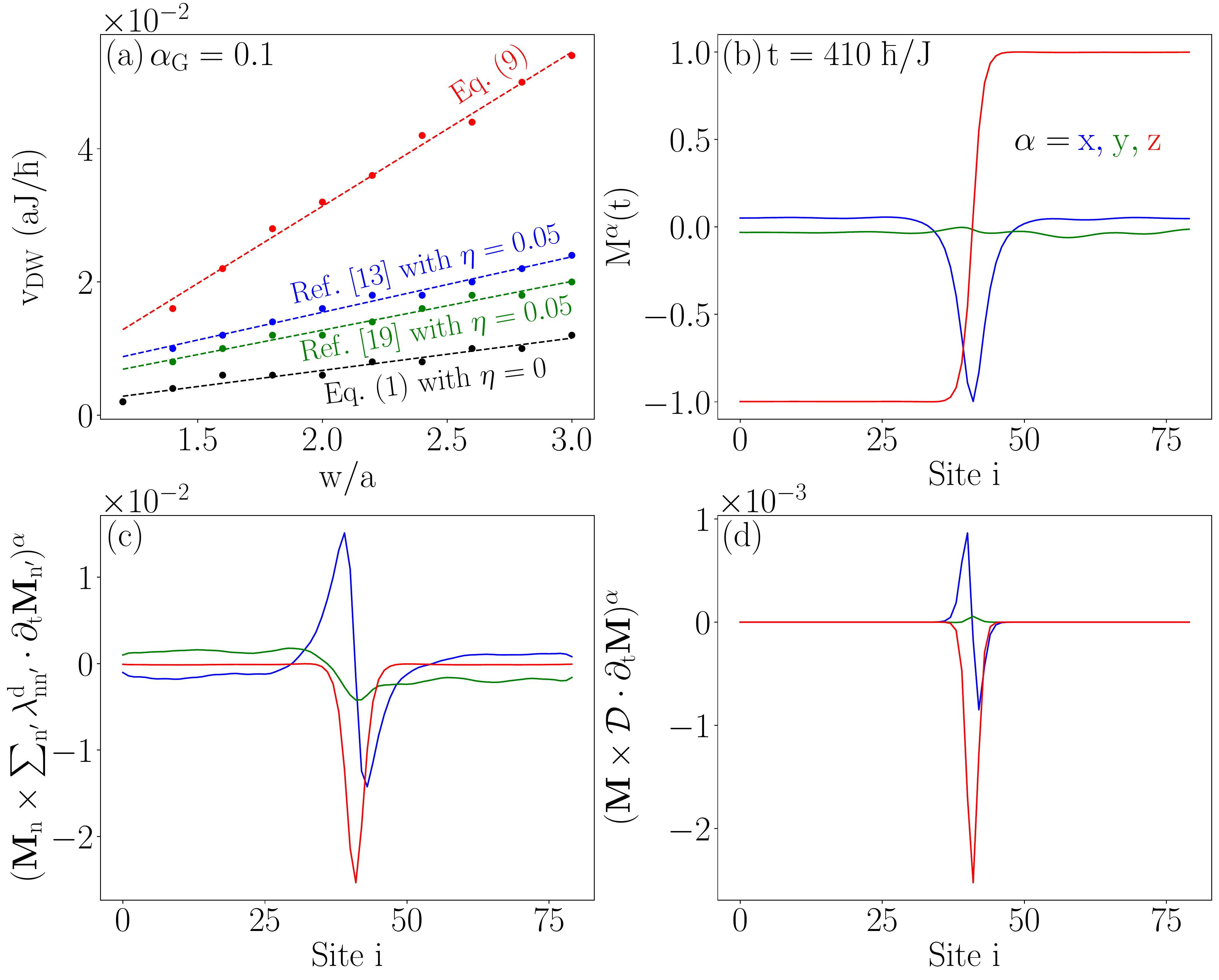}
		\caption{(a) Comparison of magnetic DW velocity $v_{\rm DW}$ vs. 
			DW width $w$ extracted from numerical simulations using: extended LLG Eq.~\eqref{eq:modifiedLLG} with SKFT-derived nonlocal damping [Eq.~\eqref{eq:damping}, red line]; extended LLG Eq.~\eqref{eq:oldLLG} with SMF-derived in Ref.~\cite{Zhang2009} nonlocal damping [Eq.~\eqref{eq:dampingTensor}, blue line] or SMF-derived nonlocal damping (green line) in Ref.~\cite{Yuan2016} [with additional term when compared to  Ref.~\cite{Zhang2009}, see Eq.~\eqref{eq:yuanLLG}]; and conventional LLG Eq.~\eqref{eq:oldLLG} with local Gilbert damping [i.e., $\eta=0$ in Eq.~\eqref{eq:dampingTensor}, black line]. (b) Spatial profile of DW within quasi-1D ferromagnetic wire at time $t=410 \ \hbar/J$, where $J$ is exchange coupling between $\mathbf{M}_n$ at NN sites, as obtained from SKFT-derived extended LLG Eq.~\eqref{eq:modifiedLLG} with nonlocal damping $\lambda^{\rm 2D}_{nn^\prime}$ 
			[Eq.~\eqref{eq:damping}]. Panels (c) and (d) plot the corresponding spatial profile of nonlocal damping across the DW in 
			(b) using SKFT-derived expression [Eqs.~\eqref{eq:modifiedLLG} and Eq.~\eqref{eq:damping}] vs.  SMF-derived~\cite{Zhang2009} expression [second term on the RHS of Eq.~\eqref{eq:dampingTensor}], respectively.}
		\label{fig:dw}
	\end{figure}
	
	The dynamics of noncollinear and noncoplanar magnetization textures, such as magnetic DWs and skyrmions, leads to pumping of charge and spin currents assumed to be captured by the spin motive force (SMF) theory~\cite{Barnes2007, Tserkovnyak2008, Duine2009}. The excess angular momentum of dynamical localized spins carried away by pumped spin current of electrons appears then as backaction torque~\cite{Petrovic2021} exerted by nonequilibrium electrons onto localized spins or, equivalently, nonlocal damping~\cite{Zhang2009, Tserkovnyak2009, Kim2012, Yuan2016}. From this viewpoint, i.e., by using expressions for pumped spin current~\cite{Zhang2009, Tserkovnyak2009, Kim2012, Yuan2016}, a particular form for nonlocal damping [second term on the RHS of Eq.~\eqref{eq:dampingTensor}] was derived in Ref.~\cite{Zhang2009} from the SMF theory, as well as extended in Ref.~\cite{Yuan2016} with an additional term, while also invoking a number of intuitively-justified but uncontrolled approximations. 
	
	In this Section, we employ an example of a magnetic field-driven DW [Fig.~\ref{fig:dw}(b)] of width $w$ within a quasi-1D ferromagnetic wire to compare its dynamics obtained by solving extended LLG Eq.~\eqref{eq:oldLLG}, which includes nonlocal damping tensor [Eq.~\eqref{eq:dampingTensor}] of Ref.~\cite{Zhang2009}, with the dynamics obtained by solving SKFT-derived extended LLG Eq.~\eqref{eq:modifiedLLG} whose nonlocal damping is different from Ref.~\cite{Zhang2009}. By neglecting nonlocal damping in Eq.~\eqref{eq:dampingTensor}, the ferromagnetic domain wall (DW) velocity $v_{\rm DW}$ is found~\cite{Tatara2019} to be directly proportional to Gilbert damping $\alpha_G$, $v_{\rm DW} \propto -{B}_{\rm ext}w\alpha_G$, assuming high external magnetic field $B_{\rm ext}$ and sufficiently small $\alpha_G$. Thus, the value of $\alpha_G$ can be extracted by measuring the DW velocity. However, experiments find that $\alpha_G$ determined in this fashion can be up to three times larger than $\alpha_G$ extracted from ferromagnetic resonance linewidth measurement scheme applied to the same material with uniform dynamical magnetization~\cite{Weindler2014}. This is considered as a strong evidence for the importance of nonlocal damping in systems hosting noncollinear magnetization textures. 
	
	In order to properly compare the effect of two different expressions for the nonlocal damping, we use $\alpha_G=0.1$ in Eq.~\eqref{eq:oldLLG} and we add the same standard local Gilbert damping term, $\alpha_G\mathbf{M}_n\times\partial_t \mathbf{M}_n$, into SKFT-derived extended LLG Eq.~\eqref{eq:modifiedLLG}. In addition, we set $\lambda_{00}^{\rm 2D} = \eta$ in Eq.~\eqref{eq:damping}, so that we can vary the same parameter $\eta$ in all versions of extended LLG Eqs.~\eqref{eq:oldLLG}, and \eqref{eq:modifiedLLG}. Note that we use $\lambda_{nn^\prime}^{\rm 2D}$ in order to include realistic decay of nonlocal damping with increasing distance $|\mathbf{r}_n-\mathbf{r}_{n^\prime}|$, thereby assuming quasi-1D wire. By changing the width of the DW, the effective damping can be extracted from the DW velocity [Fig.~\ref{fig:dw}(a)]. Figure~\ref{fig:dw}(a) shows that $v_{\rm DW}\propto w$ regardless of the specific version of nonlocal damping employed, and it increases in its presence---compare red, blue, and green data points with the black ones obtained in the absence of nonlocal damping. Nevertheless, the clear distinction between red, and blue or green data points signifies that our SKFT-derived nonlocal damping can be quite different from previously discussed SMF-derived nonlocal damping~\cite{Zhang2009, Yuan2016}, which are comparable regardless of the inclusion of the nonadiabatic terms. For example, the effective damping extracted from blue or green data points is $\mathcal{D}=0.17$ or $\mathcal{D}=0.15$, respectively, while $\lambda^{\rm 2D}_{nn^\prime}=0.48$. This distinction is further clarified by comparing spatial profiles of SKFT-derived and SMF-derived nonlocal damping in Figs.~\ref{fig:dw}(c) and~\ref{fig:dw}(d), respectively, at the instant of time used in Fig.~\ref{fig:dw}(b). In particular, the profiles differ substantially in the out-of-DW-plane or $y$-component, which is, together with the $x$-component, an order of magnitude greater in the case of SKFT-derived nonlocal damping. In addition, the SKFT-derived nonlocal damping is \textit{nonzero across the whole wire}, while the nonlocal damping in Eq.~\eqref{eq:dampingTensor} is nonzero only within the DW width, where $\mathbf{M}_n$ vectors are noncollinear [as obvious from the presence of the spatial derivative in the second term on the RHS of Eq.~\eqref{eq:dampingTensor}]. Thus, the spatial profile of SKFT-derived nonlocal damping in Fig.~\ref{fig:dw}(c) illustrates how its nonzero value in the region outside the DW width does not require noncollinearity of $\mathbf{M}_n$ vectors.
	
	Since SKFT-derived formulas are independently confirmed via numerically exact TDNEGF+LLG simulations in Figs.~\ref{fig:fig3}(c) and \ref{fig:fig3}(d), we conclude that previously derived~\cite{Zhang2009} type of nonlocal damping [second term on the RHS of Eq.~\eqref{eq:dampingTensor}] does not fully capture backaction of nonequilibrium conduction electrons onto localized spins. This could be due to nonadiabatic corrections~\cite{Tserkovnyak2008, Duine2009, Yuan2016} to spin current pumped by dynamical noncollinear magnetization textures, which are present even in the absence of disorder and SOC~\cite{Suresh2020}. One such correction was derived in Ref.~\cite{Yuan2016}, also from spin current pumping approach, thereby adding a second nonlocal damping term
	\begin{equation}\label{eq:yuanLLG}
		\eta\sum_{\beta^\prime}\Big[\left(\mathbf{M} \cdot \partial_{\beta^\prime} \partial_t \mathbf{M}\right) \mathbf{M} \times \partial_{\beta^\prime} \mathbf{M}-\mathbf{M} \times \partial_{\beta^\prime}^2 \partial_t \mathbf{M}\Big],
	\end{equation}
	into the extended LLG Eq.~\eqref{eq:oldLLG}. However, combined usage [green line in Fig.~\ref{fig:dw}(a)] of both this term and the one in Eq.~\eqref{eq:dampingTensor}  as nonlocal damping still does not match the effect of SKFT-derived nonlocal damping [compare with red line in Fig.~\ref{fig:dw}(a)] on magnetic DW. As it has been demonstrated already in Fig.~\ref{fig:pumping}, the knowledge of total spin angular momentum loss carried away by pumped spin current [Fig.~\ref{fig:pumping}(b)], as the key input in the derivations of Refs.~\cite{Zhang2009, Yuan2016}, is in general {\em insufficient} to decipher details of microscopic dynamics and dissipation exhibited by localized spins [Fig.~\ref{fig:pumping}(a)] that pump such current.

\subsection{Combining SKFT-derived nonlocal damping with first-principles calculations}\label{sec:dft}

Obtaining the closed form expressions for the nonlocal damping tensor $\lambda_{nn^\prime}$ in Secs.~\ref{sec:noSoc} and \ref{sec:soc} was made possible by using simplistic model Hamiltonians and geometries. For realistic materials and more complicated geometries, we provide in this Section general formulas which can be combined with DFT quantities and evaluated numerically.

Notably, the time-retarded dissipation kernel in Eq.~\eqref{eq:dissipationKernels}, from which $\lambda_{nn^\prime}$ is extracted, depends on the Keldysh GFs. The same GFs are also commonly used in first-principles calculations of conventional Gilbert damping scalar parameter via Kubo-type formulas~\cite{Gilmore2007, Ebert2011, Mankovsky2013,Hou2019,Guimaraes2019}. Specifically, the retarded/advanced GFs are obtained from first-principles Hamiltonians $\hat H^{\rm DFT}$ DFT as \mbox{$\hat G^{R/A}(\varepsilon)=\big[ \varepsilon - \hat H^{\rm DFT} + \hat\Sigma^{R/A}(\epsilon) \big]^{-1}$}. Here, $\hat\Sigma^{R/A}(\varepsilon)$ are the retarded/advanced self-energies~\cite{Ryndyk2016, Nunez2008} describing escape rate of electrons into NM leads, allowing for open-system setups akin to the scattering theory-derived formula for Gilbert damping~\cite{Brataas2008,Brataas2011} and its computational implementation with DFT Hamiltonians~\cite{Starikov2010,Starikov2018}. Since escape rates are encoded by imaginary part of the self-energy, such calculations do not require $i \eta$ imaginary parameter introduced by hand when using  Kubo-type formulas~\cite{Gilmore2007, Ebert2011, Mankovsky2013,Hou2019,Guimaraes2019} (where $\eta \rightarrow 0$ leads to unphysical divergent results~\cite{Costa2015,Edwards2016,Mahfouzi2017a}). Therefore, $\hat H^{\rm DFT}$ can be used as an input to compute the nonlocal damping tensor, via the calculation of the GFs $\hat G^{R/A}(\varepsilon)$ and the spectral function $\hat A(\varepsilon) = i\big[\hat G^R(\varepsilon) -\hat G^A(\varepsilon)\big]$.

For these purposes, it is convenient to separate the nonlocal damping tensor into its symmetric and antisymmetric components, $\lambda_{nn^\prime}^{\alpha\beta}=\lambda_{nn^\prime}^{(\alpha\beta)} + \lambda_{nn^\prime}^{[\alpha\beta]}$, where the parenthesis (brackets) indicate that surrounded indices have been (anti)symmetrized. They are given by
\begin{subequations}\label{eq:formulasfordft}
\begin{align}
\lambda^{(\alpha\beta)}_{nn^\prime} &=  - \frac{J_{sd}^2}{2\pi} \int\!d\varepsilon \, \frac{\partial f}{\partial \varepsilon} \mathrm{Tr_{spin}} \left [\sigma^\alpha A_{nn^\prime} \sigma^\beta A_{n^\prime n}\right],  \label{eq:dftsymm} \\
 \lambda_{nn^\prime}^{[\alpha\beta]} &= -\frac{2J_{sd}^2}{\pi}\int\!d\varepsilon \, \frac{\partial f}{\partial\varepsilon} {\rm Tr_{spin}}\big[\sigma^\alpha{\rm Re}\,\hat G^R_{nn^\prime}\sigma^\beta A_{n^\prime n} \nonumber\\
        &-\sigma^\alpha A_{nn^\prime}\sigma^\beta {\rm Re}\,\hat G^R_{n^\prime n} \big] + \frac{J_{sd}^2}{2\pi}\int\!d\varepsilon \, (1-2f) \nonumber \\
        &\times{\rm Tr_{spin}}\big[\sigma^\alpha{\rm Re}\,\hat G^R_{nn^\prime}\sigma^\beta \frac{\partial A_{n^\prime n}}{\partial\varepsilon} -\sigma^\alpha \frac{\partial A_{nn^\prime }}{\partial\varepsilon}\sigma^\beta {\rm Re} \, \hat G^R_{n^\prime n} \big], \label{eq:dftantisymm}
\end{align}
\end{subequations}
where $f(\varepsilon)$ is the Fermi function, and the trace is taken in the spin space. The antisymmetric component either vanishes in the presence of inversion symmetry, or is often orders of magnitude smaller than the symmetric one. Therefore, it is absent in our results for simple models on hypercubic lattices. As such, the nonlocal damping tensors in Eqs.~\eqref{eq:damping} and \eqref{eq:dampingSOC}, are fully symmetric and special case of Eq.~\eqref{eq:dftsymm} when  considering specific energy-momentum dispersions and assuming zero temperature.

	\section{Conclusions and Outlook}\label{sec:conclusions}
	
	In conclusion, we derived a novel formula, displayed as Eq.~\eqref{eq:formulasfordft}, for magnetization damping of a metallic ferromagnet via unexploited for this purpose rigorous approach offered by the Schwinger-Keldysh nonequilibrium field theory~\cite{Kamenev2011}. Our formulas could open a new route for calculations of Gilbert damping of realistic materials by employing first-principles Hamiltonian $\hat H^{\rm DFT}$ from density functional theory (DFT) as an input, as discussed in Sec.~\ref{sec:dft}. Although a thorough numerical exploration of a small two-spin system based on SKFT was recently pursued in Ref.~\cite{Leiva2023}, our Eqs.~\eqref{eq:formulasfordft} are not only applicable for large systems of many localized spins, but are also refined into readily computable expressions that depend on accessible quantities.
 
    While traditional, Kubo linear-response~\cite{Kambersky1976,Kambersky1984,Kambersky2007,Ebert2011,Thonig2014} or scattering theory~\cite{Brataas2008} based derivations  produce spatially uniform scalar $\alpha_G$, SKFT-derived damping in Eqs.~\eqref{eq:formulasfordft} is intrinsically nonlocal and nonuniform as it depends on the coordinates of local magnetization at two points in space $\mathbf{r}_n$ and $\mathbf{r}_{n^\prime}$. In the cases of model Hamiltonians in 1D--3D, we reduced Eqs.~\eqref{eq:formulasfordft} to analytical expressions for magnetization damping [Eq.~\eqref{eq:damping}], thereby making it possible to understand the consequences of such fundamental nonlocality and nonuniformity on local magnetization dynamics, such as: ({\em i}) damping in Eq.~\eqref{eq:damping} osc illates with the distance between $\mathbf{x}$ and $\mathbf{x^\prime}$ where the period of such oscillation is governed by the Fermi wavevector $k_F$ [Figs.~\ref{fig:fig1}(c), \ref{fig:fig3}(c), and \ref{fig:fig3}(d)]; ({\em ii}) it always leads to nonuniform local magnetization dynamics [Fig.~\ref{fig:pumping}(a)], even though spin pumping from it can appear  [Fig.~\ref{fig:pumping}(b)] as if it is driven by usually analyzed~\cite{Brataas2008,Tserkovnyak2005} uniform local magnetization (or, equivalently, macrospin); (\textit{iii}) when applied to noncollinear magnetic textures, such as DWs, it produces an order of magnitude larger damping and, therefore, DW wall velocity, than predicted by previously derived~\cite{Zhang2009} nonlocal damping [second term on the RHS of Eq.~\eqref{eq:dampingTensor}]. Remarkably, solutions of SKFT-based extended LLG Eq.~\eqref{eq:modifiedLLG} are fully corroborated by numerically exact TDNEGF+LLG simulations~\cite{Bajpai2019,Petrovic2018,Petrovic2021,Suresh2020} in 1D, despite the fact that several approximations are employed in SKFT-based derivations. Finally, while conventional understanding of the origin of Gilbert damping scalar parameter $\alpha_G$ requires SOC to be nonzero~\cite{Hickey2009,Mondal2017}, our nonlocal damping is nonzero [Eq.~\eqref{eq:damping}] even in the absence of SOC due to inevitable delay~\cite{Sayad2015,Bajpai2019} in electronic spin responding  to motion of localized classical spins. For typical values of \mbox{$J_{sd}\sim 0.1$ eV}~\cite{Cooper1967} and NN hopping parameter $\gamma\sim 1$ eV, the magnitude of nonlocal damping is \mbox{$\lambda_{nn^\prime}^D\lesssim 0.01$}, relevant even in metallic magnets with conventional local damping \mbox{$\alpha_G\sim 0.01$~\cite{Weindler2014}}. By  switching SOC on, such nonlocal damping becomes additionally anisotropic [Eq.~\eqref{eq:dampingSOC}].  
	
	\begin{acknowledgements}
		This work was supported by the US National Science Foundation (NSF) Grant No. ECCS 1922689.
	\end{acknowledgements}

\bibliography{master-biblio}

\begin{thebibliography}{76}%
\makeatletter
\providecommand \@ifxundefined [1]{%
 \@ifx{#1\undefined}
}%
\providecommand \@ifnum [1]{%
 \ifnum #1\expandafter \@firstoftwo
 \else \expandafter \@secondoftwo
 \fi
}%
\providecommand \@ifx [1]{%
 \ifx #1\expandafter \@firstoftwo
 \else \expandafter \@secondoftwo
 \fi
}%
\providecommand \natexlab [1]{#1}%
\providecommand \enquote  [1]{``#1''}%
\providecommand \bibnamefont  [1]{#1}%
\providecommand \bibfnamefont [1]{#1}%
\providecommand \citenamefont [1]{#1}%
\providecommand \href@noop [0]{\@secondoftwo}%
\providecommand \href [0]{\begingroup \@sanitize@url \@href}%
\providecommand \@href[1]{\@@startlink{#1}\@@href}%
\providecommand \@@href[1]{\endgroup#1\@@endlink}%
\providecommand \@sanitize@url [0]{\catcode `\\12\catcode `\$12\catcode
  `\&12\catcode `\#12\catcode `\^12\catcode `\_12\catcode `\%12\relax}%
\providecommand \@@startlink[1]{}%
\providecommand \@@endlink[0]{}%
\providecommand \url  [0]{\begingroup\@sanitize@url \@url }%
\providecommand \@url [1]{\endgroup\@href {#1}{\urlprefix }}%
\providecommand \urlprefix  [0]{URL }%
\providecommand \Eprint [0]{\href }%
\providecommand \doibase [0]{https://doi.org/}%
\providecommand \selectlanguage [0]{\@gobble}%
\providecommand \bibinfo  [0]{\@secondoftwo}%
\providecommand \bibfield  [0]{\@secondoftwo}%
\providecommand \translation [1]{[#1]}%
\providecommand \BibitemOpen [0]{}%
\providecommand \bibitemStop [0]{}%
\providecommand \bibitemNoStop [0]{.\EOS\space}%
\providecommand \EOS [0]{\spacefactor3000\relax}%
\providecommand \BibitemShut  [1]{\csname bibitem#1\endcsname}%
\let\auto@bib@innerbib\@empty
\bibitem [{\citenamefont {Landau}\ and\ \citenamefont
  {Lifshitz}(1935)}]{Landau1935}%
  \BibitemOpen
  \bibfield  {author} {\bibinfo {author} {\bibfnamefont {L.~D.}\ \bibnamefont
  {Landau}}\ and\ \bibinfo {author} {\bibfnamefont {E.~M.}\ \bibnamefont
  {Lifshitz}},\ }\bibfield  {title} {\bibinfo {title} {On the theory of the
  dispersion of magnetic permeability in ferromagnetic bodies},\ }\href@noop {}
  {\bibfield  {journal} {\bibinfo  {journal} {Phys. Z. Sowjetunion}\ }\textbf
  {\bibinfo {volume} {8}},\ \bibinfo {pages} {153} (\bibinfo {year}
  {1935})}\BibitemShut {NoStop}%
\bibitem [{\citenamefont {Berkov}\ and\ \citenamefont
  {Miltat}(2008)}]{Berkov2008}%
  \BibitemOpen
  \bibfield  {author} {\bibinfo {author} {\bibfnamefont {D.~V.}\ \bibnamefont
  {Berkov}}\ and\ \bibinfo {author} {\bibfnamefont {J.}~\bibnamefont
  {Miltat}},\ }\bibfield  {title} {\bibinfo {title} {Spin-torque driven
  magnetization dynamics: Micromagnetic modeling},\ }\href
  {https://doi.org/10.1016/j.jmmm.2007.12.023} {\bibfield  {journal} {\bibinfo
  {journal} {J. Magn. Magn. Mater.}\ }\textbf {\bibinfo {volume} {320}},\
  \bibinfo {pages} {1238} (\bibinfo {year} {2008})}\BibitemShut {NoStop}%
\bibitem [{\citenamefont {Kim}(2010)}]{Kim2010}%
  \BibitemOpen
  \bibfield  {author} {\bibinfo {author} {\bibfnamefont {S.-K.}\ \bibnamefont
  {Kim}},\ }\bibfield  {title} {\bibinfo {title} {Micromagnetic computer
  simulations of spin waves in nanometre-scale patterned magnetic elements},\
  }\href {https://doi.org/10.1088/0022-3727/43/26/264004} {\bibfield  {journal}
  {\bibinfo  {journal} {J. Phys. D: Appl. Phys.}\ }\textbf {\bibinfo {volume}
  {43}},\ \bibinfo {pages} {264004} (\bibinfo {year} {2010})}\BibitemShut
  {NoStop}%
\bibitem [{\citenamefont {Evans}\ \emph {et~al.}(2014)\citenamefont {Evans},
  \citenamefont {Fan}, \citenamefont {Chureemart}, \citenamefont {Ostler},
  \citenamefont {Ellis},\ and\ \citenamefont {Chantrell}}]{Evans2014}%
  \BibitemOpen
  \bibfield  {author} {\bibinfo {author} {\bibfnamefont {R.}~\bibnamefont
  {Evans}}, \bibinfo {author} {\bibfnamefont {W.}~\bibnamefont {Fan}}, \bibinfo
  {author} {\bibfnamefont {P.}~\bibnamefont {Chureemart}}, \bibinfo {author}
  {\bibfnamefont {T.}~\bibnamefont {Ostler}}, \bibinfo {author} {\bibfnamefont
  {M.~O.}\ \bibnamefont {Ellis}},\ and\ \bibinfo {author} {\bibfnamefont
  {R.}~\bibnamefont {Chantrell}},\ }\bibfield  {title} {\bibinfo {title}
  {Atomistic spin model simulations of magnetic nanomaterials},\ }\href
  {https://doi.org/10.1088/0953-8984/26/10/103202} {\bibfield  {journal}
  {\bibinfo  {journal} {J. Phys.: Condens. Matter}\ }\textbf {\bibinfo {volume}
  {26}},\ \bibinfo {pages} {103202} (\bibinfo {year} {2014})}\BibitemShut
  {NoStop}%
\bibitem [{\citenamefont {Garc\'ia-Ga\'itan}\ and\ \citenamefont
  {Nikoli\'{c}}()}]{GarciaGaitan2023}%
  \BibitemOpen
  \bibfield  {author} {\bibinfo {author} {\bibfnamefont {F.}~\bibnamefont
  {Garc\'ia-Ga\'itan}}\ and\ \bibinfo {author} {\bibfnamefont {B.~K.}\
  \bibnamefont {Nikoli\'{c}}},\ }\href@noop {} {\bibinfo {title} {Fate of
  entanglement in magnetism under {Lindbladian} or non-{Markovian} dynamics and
  conditions for their transition to {Landau}-{Lifshitz}-{Gilbert} classical
  dynamics}},\ \Eprint {https://arxiv.org/abs/2303.17596 (2023)}
  {arXiv:2303.17596 (2023)} \BibitemShut {NoStop}%
\bibitem [{\citenamefont {Saslow}(2009)}]{Saslow2009}%
  \BibitemOpen
  \bibfield  {author} {\bibinfo {author} {\bibfnamefont {W.~M.}\ \bibnamefont
  {Saslow}},\ }\bibfield  {title} {\bibinfo {title} {{L}andau-{L}ifshitz or
  {G}ilbert damping? that is the question},\ }\href
  {https://doi.org/10.1063/1.3077204} {\bibfield  {journal} {\bibinfo
  {journal} {J. Appl. Phys.}\ }\textbf {\bibinfo {volume} {105}},\ \bibinfo
  {pages} {07D315} (\bibinfo {year} {2009})}\BibitemShut {NoStop}%
\bibitem [{\citenamefont {Gilbert}(2004)}]{Gilbert2004}%
  \BibitemOpen
  \bibfield  {author} {\bibinfo {author} {\bibfnamefont {T.}~\bibnamefont
  {Gilbert}},\ }\bibfield  {title} {\bibinfo {title} {A phenomenological theory
  of damping in ferromagnetic materials},\ }\href
  {https://doi.org/10.1109/TMAG.2004.836740} {\bibfield  {journal} {\bibinfo
  {journal} {{IEEE} Trans. Magn.}\ }\textbf {\bibinfo {volume} {40}},\ \bibinfo
  {pages} {3443} (\bibinfo {year} {2004})}\BibitemShut {NoStop}%
\bibitem [{\citenamefont {Brataas}\ \emph {et~al.}(2008)\citenamefont
  {Brataas}, \citenamefont {Tserkovnyak},\ and\ \citenamefont
  {Bauer}}]{Brataas2008}%
  \BibitemOpen
  \bibfield  {author} {\bibinfo {author} {\bibfnamefont {A.}~\bibnamefont
  {Brataas}}, \bibinfo {author} {\bibfnamefont {Y.}~\bibnamefont
  {Tserkovnyak}},\ and\ \bibinfo {author} {\bibfnamefont {G.}~\bibnamefont
  {Bauer}},\ }\bibfield  {title} {\bibinfo {title} {Scattering theory of
  {Gilbert} damping},\ }\href {https://doi.org/10.1103/PHYSREVLETT.101.037207}
  {\bibfield  {journal} {\bibinfo  {journal} {Phys. Rev. Lett.}\ }\textbf
  {\bibinfo {volume} {101}},\ \bibinfo {pages} {037207} (\bibinfo {year}
  {2008})}\BibitemShut {NoStop}%
\bibitem [{\citenamefont {Thonig}\ and\ \citenamefont
  {Henk}(2014)}]{Thonig2014}%
  \BibitemOpen
  \bibfield  {author} {\bibinfo {author} {\bibfnamefont {D.}~\bibnamefont
  {Thonig}}\ and\ \bibinfo {author} {\bibfnamefont {J.}~\bibnamefont {Henk}},\
  }\bibfield  {title} {\bibinfo {title} {{Gilbert} damping tensor within the
  breathing {Fermi} surface model: anisotropy and non-locality},\ }\href
  {https://doi.org/10.1088/1367-2630/16/1/013032} {\bibfield  {journal}
  {\bibinfo  {journal} {New J. Phys.}\ }\textbf {\bibinfo {volume} {16}},\
  \bibinfo {pages} {013032} (\bibinfo {year} {2014})}\BibitemShut {NoStop}%
\bibitem [{\citenamefont {Weindler}\ \emph {et~al.}(2014)\citenamefont
  {Weindler}, \citenamefont {Bauer}, \citenamefont {Islinger}, \citenamefont
  {Boehm}, \citenamefont {Chauleau},\ and\ \citenamefont
  {Back}}]{Weindler2014}%
  \BibitemOpen
  \bibfield  {author} {\bibinfo {author} {\bibfnamefont {T.}~\bibnamefont
  {Weindler}}, \bibinfo {author} {\bibfnamefont {H.~G.}\ \bibnamefont {Bauer}},
  \bibinfo {author} {\bibfnamefont {R.}~\bibnamefont {Islinger}}, \bibinfo
  {author} {\bibfnamefont {B.}~\bibnamefont {Boehm}}, \bibinfo {author}
  {\bibfnamefont {J.-Y.}\ \bibnamefont {Chauleau}},\ and\ \bibinfo {author}
  {\bibfnamefont {C.~H.}\ \bibnamefont {Back}},\ }\bibfield  {title} {\bibinfo
  {title} {Magnetic damping: Domain wall dynamics versus local ferromagnetic
  resonance},\ }\href {https://doi.org/10.1103/PhysRevLett.113.237204}
  {\bibfield  {journal} {\bibinfo  {journal} {Phys. Rev. Lett.}\ }\textbf
  {\bibinfo {volume} {113}},\ \bibinfo {pages} {237204} (\bibinfo {year}
  {2014})}\BibitemShut {NoStop}%
\bibitem [{\citenamefont {Soumah}\ \emph {et~al.}(2018)\citenamefont {Soumah},
  \citenamefont {Beaulieu}, \citenamefont {Qassym}, \citenamefont
  {Carr{\'{e}}t{\'{e}}ro}, \citenamefont {Jacquet}, \citenamefont
  {Lebourgeois}, \citenamefont {Youssef}, \citenamefont {Bortolotti},
  \citenamefont {Cros},\ and\ \citenamefont {Anane}}]{Soumah2018}%
  \BibitemOpen
  \bibfield  {author} {\bibinfo {author} {\bibfnamefont {L.}~\bibnamefont
  {Soumah}}, \bibinfo {author} {\bibfnamefont {N.}~\bibnamefont {Beaulieu}},
  \bibinfo {author} {\bibfnamefont {L.}~\bibnamefont {Qassym}}, \bibinfo
  {author} {\bibfnamefont {C.}~\bibnamefont {Carr{\'{e}}t{\'{e}}ro}}, \bibinfo
  {author} {\bibfnamefont {E.}~\bibnamefont {Jacquet}}, \bibinfo {author}
  {\bibfnamefont {R.}~\bibnamefont {Lebourgeois}}, \bibinfo {author}
  {\bibfnamefont {J.~B.}\ \bibnamefont {Youssef}}, \bibinfo {author}
  {\bibfnamefont {P.}~\bibnamefont {Bortolotti}}, \bibinfo {author}
  {\bibfnamefont {V.}~\bibnamefont {Cros}},\ and\ \bibinfo {author}
  {\bibfnamefont {A.}~\bibnamefont {Anane}},\ }\bibfield  {title} {\bibinfo
  {title} {Ultra-low damping insulating magnetic thin films get
  perpendicular},\ }\href {https://doi.org/10.1038/s41467-018-05732-1}
  {\bibfield  {journal} {\bibinfo  {journal} {Nat. Commun.}\ }\textbf {\bibinfo
  {volume} {9}},\ \bibinfo {pages} {3355} (\bibinfo {year} {2018})}\BibitemShut
  {NoStop}%
\bibitem [{\citenamefont {Schoen}\ \emph {et~al.}(2016)\citenamefont {Schoen},
  \citenamefont {Thonig}, \citenamefont {Schneider}, \citenamefont {Silva},
  \citenamefont {Nembach}, \citenamefont {Eriksson}, \citenamefont {Karis},\
  and\ \citenamefont {Shaw}}]{Schoen2016}%
  \BibitemOpen
  \bibfield  {author} {\bibinfo {author} {\bibfnamefont {M.~A.~W.}\
  \bibnamefont {Schoen}}, \bibinfo {author} {\bibfnamefont {D.}~\bibnamefont
  {Thonig}}, \bibinfo {author} {\bibfnamefont {M.~L.}\ \bibnamefont
  {Schneider}}, \bibinfo {author} {\bibfnamefont {T.~J.}\ \bibnamefont
  {Silva}}, \bibinfo {author} {\bibfnamefont {H.~T.}\ \bibnamefont {Nembach}},
  \bibinfo {author} {\bibfnamefont {O.}~\bibnamefont {Eriksson}}, \bibinfo
  {author} {\bibfnamefont {O.}~\bibnamefont {Karis}},\ and\ \bibinfo {author}
  {\bibfnamefont {J.~M.}\ \bibnamefont {Shaw}},\ }\bibfield  {title} {\bibinfo
  {title} {Ultra-low magnetic damping of a metallic ferromagnet},\ }\href
  {https://doi.org/10.1038/nphys3770} {\bibfield  {journal} {\bibinfo
  {journal} {Nat. Phys.}\ }\textbf {\bibinfo {volume} {12}},\ \bibinfo {pages}
  {839} (\bibinfo {year} {2016})}\BibitemShut {NoStop}%
\bibitem [{\citenamefont {Zhang}\ and\ \citenamefont
  {Zhang}(2009)}]{Zhang2009}%
  \BibitemOpen
  \bibfield  {author} {\bibinfo {author} {\bibfnamefont {S.}~\bibnamefont
  {Zhang}}\ and\ \bibinfo {author} {\bibfnamefont {S.~L.}\ \bibnamefont
  {Zhang}},\ }\bibfield  {title} {\bibinfo {title} {Generalization of the
  {Landau}-{Lifshitz}-{Gilbert} equation for conducting ferromagnets},\ }\href
  {https://doi.org/10.1103/PHYSREVLETT.102.086601} {\bibfield  {journal}
  {\bibinfo  {journal} {Phys. Rev. Lett.}\ }\textbf {\bibinfo {volume} {102}},\
  \bibinfo {pages} {086601} (\bibinfo {year} {2009})}\BibitemShut {NoStop}%
\bibitem [{\citenamefont {Foros}\ \emph {et~al.}(2008)\citenamefont {Foros},
  \citenamefont {Brataas}, \citenamefont {Tserkovnyak},\ and\ \citenamefont
  {Bauer}}]{Foros2008}%
  \BibitemOpen
  \bibfield  {author} {\bibinfo {author} {\bibfnamefont {J.}~\bibnamefont
  {Foros}}, \bibinfo {author} {\bibfnamefont {A.}~\bibnamefont {Brataas}},
  \bibinfo {author} {\bibfnamefont {Y.}~\bibnamefont {Tserkovnyak}},\ and\
  \bibinfo {author} {\bibfnamefont {G.}~\bibnamefont {Bauer}},\ }\bibfield
  {title} {\bibinfo {title} {Current-induced noise and damping in nonuniform
  ferromagnets},\ }\href {https://doi.org/10.1103/PhysRevB.78.140402}
  {\bibfield  {journal} {\bibinfo  {journal} {Phys. Rev. B}\ }\textbf {\bibinfo
  {volume} {78}},\ \bibinfo {pages} {140402} (\bibinfo {year}
  {2008})}\BibitemShut {NoStop}%
\bibitem [{\citenamefont {Hankiewicz}\ \emph {et~al.}(2008)\citenamefont
  {Hankiewicz}, \citenamefont {Vignale},\ and\ \citenamefont
  {Tserkovnyak}}]{Hankiewicz2008}%
  \BibitemOpen
  \bibfield  {author} {\bibinfo {author} {\bibfnamefont {E.}~\bibnamefont
  {Hankiewicz}}, \bibinfo {author} {\bibfnamefont {G.}~\bibnamefont
  {Vignale}},\ and\ \bibinfo {author} {\bibfnamefont {Y.}~\bibnamefont
  {Tserkovnyak}},\ }\bibfield  {title} {\bibinfo {title} {Inhomogeneous
  {Gilbert} damping from impurities and electron-electron interactions},\
  }\href {https://doi.org/10.1103/PhysRevB.78.020404} {\bibfield  {journal}
  {\bibinfo  {journal} {Phys. Rev. B}\ }\textbf {\bibinfo {volume} {78}},\
  \bibinfo {pages} {020404(R)} (\bibinfo {year} {2008})}\BibitemShut {NoStop}%
\bibitem [{\citenamefont {Tserkovnyak}\ and\ \citenamefont
  {Mecklenburg}(2008)}]{Tserkovnyak2008}%
  \BibitemOpen
  \bibfield  {author} {\bibinfo {author} {\bibfnamefont {Y.}~\bibnamefont
  {Tserkovnyak}}\ and\ \bibinfo {author} {\bibfnamefont {M.}~\bibnamefont
  {Mecklenburg}},\ }\bibfield  {title} {\bibinfo {title} {Electron transport
  driven by nonequilibrium magnetic textures},\ }\href
  {https://doi.org/10.1103/PhysRevB.77.134407} {\bibfield  {journal} {\bibinfo
  {journal} {Phys. Rev. B}\ }\textbf {\bibinfo {volume} {77}},\ \bibinfo
  {pages} {134407} (\bibinfo {year} {2008})}\BibitemShut {NoStop}%
\bibitem [{\citenamefont {Tserkovnyak}\ \emph {et~al.}(2009)\citenamefont
  {Tserkovnyak}, \citenamefont {Hankiewicz},\ and\ \citenamefont
  {Vignale}}]{Tserkovnyak2009}%
  \BibitemOpen
  \bibfield  {author} {\bibinfo {author} {\bibfnamefont {Y.}~\bibnamefont
  {Tserkovnyak}}, \bibinfo {author} {\bibfnamefont {E.~M.}\ \bibnamefont
  {Hankiewicz}},\ and\ \bibinfo {author} {\bibfnamefont {G.}~\bibnamefont
  {Vignale}},\ }\bibfield  {title} {\bibinfo {title} {Transverse spin diffusion
  in ferromagnets},\ }\href {https://doi.org/10.1103/physrevb.79.094415}
  {\bibfield  {journal} {\bibinfo  {journal} {Phys. Rev. B}\ }\textbf {\bibinfo
  {volume} {79}},\ \bibinfo {pages} {094415} (\bibinfo {year}
  {2009})}\BibitemShut {NoStop}%
\bibitem [{\citenamefont {Kim}\ \emph {et~al.}(2012)\citenamefont {Kim},
  \citenamefont {Moon}, \citenamefont {Lee},\ and\ \citenamefont
  {Lee}}]{Kim2012}%
  \BibitemOpen
  \bibfield  {author} {\bibinfo {author} {\bibfnamefont {K.-W.}\ \bibnamefont
  {Kim}}, \bibinfo {author} {\bibfnamefont {J.-H.}\ \bibnamefont {Moon}},
  \bibinfo {author} {\bibfnamefont {K.-J.}\ \bibnamefont {Lee}},\ and\ \bibinfo
  {author} {\bibfnamefont {H.-W.}\ \bibnamefont {Lee}},\ }\bibfield  {title}
  {\bibinfo {title} {Prediction of giant spin motive force due to {Rashba}
  spin-orbit coupling},\ }\href
  {https://doi.org/10.1103/PhysRevLett.108.217202} {\bibfield  {journal}
  {\bibinfo  {journal} {Phys. Rev. Lett.}\ }\textbf {\bibinfo {volume} {108}},\
  \bibinfo {pages} {217202} (\bibinfo {year} {2012})}\BibitemShut {NoStop}%
\bibitem [{\citenamefont {Yuan}\ \emph {et~al.}(2016)\citenamefont {Yuan},
  \citenamefont {Yuan}, \citenamefont {Xia},\ and\ \citenamefont
  {Wang}}]{Yuan2016}%
  \BibitemOpen
  \bibfield  {author} {\bibinfo {author} {\bibfnamefont {H.}~\bibnamefont
  {Yuan}}, \bibinfo {author} {\bibfnamefont {Z.}~\bibnamefont {Yuan}}, \bibinfo
  {author} {\bibfnamefont {K.}~\bibnamefont {Xia}},\ and\ \bibinfo {author}
  {\bibfnamefont {X.~R.}\ \bibnamefont {Wang}},\ }\bibfield  {title} {\bibinfo
  {title} {Influence of nonlocal damping on the field-driven domain wall
  motion},\ }\href {https://doi.org/10.1103/PhysRevB.94.064415} {\bibfield
  {journal} {\bibinfo  {journal} {Phys. Rev. B}\ }\textbf {\bibinfo {volume}
  {94}},\ \bibinfo {pages} {064415} (\bibinfo {year} {2016})}\BibitemShut
  {NoStop}%
\bibitem [{\citenamefont {Verba}\ \emph {et~al.}(2018)\citenamefont {Verba},
  \citenamefont {Tiberkevich},\ and\ \citenamefont {Slavin}}]{Verba2018}%
  \BibitemOpen
  \bibfield  {author} {\bibinfo {author} {\bibfnamefont {R.}~\bibnamefont
  {Verba}}, \bibinfo {author} {\bibfnamefont {V.}~\bibnamefont {Tiberkevich}},\
  and\ \bibinfo {author} {\bibfnamefont {A.}~\bibnamefont {Slavin}},\
  }\bibfield  {title} {\bibinfo {title} {Damping of linear spin-wave modes in
  magnetic nanostructures: Local, nonlocal, and coordinate-dependent damping},\
  }\href {https://doi.org/10.1103/PHYSREVB.98.104408} {\bibfield  {journal}
  {\bibinfo  {journal} {Phys. Rev. B}\ }\textbf {\bibinfo {volume} {98}},\
  \bibinfo {pages} {104408} (\bibinfo {year} {2018})}\BibitemShut {NoStop}%
\bibitem [{\citenamefont {Mankovsky}\ \emph {et~al.}(2018)\citenamefont
  {Mankovsky}, \citenamefont {Wimmer},\ and\ \citenamefont
  {Ebert}}]{Mankovsky2018}%
  \BibitemOpen
  \bibfield  {author} {\bibinfo {author} {\bibfnamefont {S.}~\bibnamefont
  {Mankovsky}}, \bibinfo {author} {\bibfnamefont {S.}~\bibnamefont {Wimmer}},\
  and\ \bibinfo {author} {\bibfnamefont {H.}~\bibnamefont {Ebert}},\ }\bibfield
   {title} {\bibinfo {title} {Gilbert damping in noncollinear magnetic
  systems},\ }\href {https://doi.org/10.1103/PhysRevB.98.104406} {\bibfield
  {journal} {\bibinfo  {journal} {Phys. Rev. B}\ }\textbf {\bibinfo {volume}
  {98}},\ \bibinfo {pages} {104406} (\bibinfo {year} {2018})}\BibitemShut
  {NoStop}%
\bibitem [{\citenamefont {Mondal}\ \emph {et~al.}(2017)\citenamefont {Mondal},
  \citenamefont {Berritta}, \citenamefont {Nandy},\ and\ \citenamefont
  {Oppeneer}}]{Mondal2017}%
  \BibitemOpen
  \bibfield  {author} {\bibinfo {author} {\bibfnamefont {R.}~\bibnamefont
  {Mondal}}, \bibinfo {author} {\bibfnamefont {M.}~\bibnamefont {Berritta}},
  \bibinfo {author} {\bibfnamefont {A.}~\bibnamefont {Nandy}},\ and\ \bibinfo
  {author} {\bibfnamefont {P.}~\bibnamefont {Oppeneer}},\ }\bibfield  {title}
  {\bibinfo {title} {Relativistic theory of magnetic inertia in ultrafast spin
  dynamics},\ }\href {https://doi.org/10.1103/PhysRevB.96.024425} {\bibfield
  {journal} {\bibinfo  {journal} {Phys. Rev. B}\ }\textbf {\bibinfo {volume}
  {96}},\ \bibinfo {pages} {024425} (\bibinfo {year} {2017})}\BibitemShut
  {NoStop}%
\bibitem [{\citenamefont {Hickey}\ and\ \citenamefont
  {Moodera}(2009)}]{Hickey2009}%
  \BibitemOpen
  \bibfield  {author} {\bibinfo {author} {\bibfnamefont {M.~C.}\ \bibnamefont
  {Hickey}}\ and\ \bibinfo {author} {\bibfnamefont {J.~S.}\ \bibnamefont
  {Moodera}},\ }\bibfield  {title} {\bibinfo {title} {Origin of intrinsic
  {Gilbert} damping},\ }\href {https://doi.org/10.1103/PhysRevLett.102.137601}
  {\bibfield  {journal} {\bibinfo  {journal} {Phys. Rev. Lett.}\ }\textbf
  {\bibinfo {volume} {102}},\ \bibinfo {pages} {137601} (\bibinfo {year}
  {2009})}\BibitemShut {NoStop}%
\bibitem [{\citenamefont {Starikov}\ \emph {et~al.}(2010)\citenamefont
  {Starikov}, \citenamefont {Kelly}, \citenamefont {Brataas}, \citenamefont
  {Tserkovnyak},\ and\ \citenamefont {Bauer}}]{Starikov2010}%
  \BibitemOpen
  \bibfield  {author} {\bibinfo {author} {\bibfnamefont {A.~A.}\ \bibnamefont
  {Starikov}}, \bibinfo {author} {\bibfnamefont {P.~J.}\ \bibnamefont {Kelly}},
  \bibinfo {author} {\bibfnamefont {A.}~\bibnamefont {Brataas}}, \bibinfo
  {author} {\bibfnamefont {Y.}~\bibnamefont {Tserkovnyak}},\ and\ \bibinfo
  {author} {\bibfnamefont {G.~E.~W.}\ \bibnamefont {Bauer}},\ }\bibfield
  {title} {\bibinfo {title} {Unified first-principles study of {Gilbert}
  damping, spin-flip diffusion, and resistivity in transition metal alloys},\
  }\href {https://doi.org/10.1103/PhysRevLett.105.236601} {\bibfield  {journal}
  {\bibinfo  {journal} {Phys. Rev. Lett.}\ }\textbf {\bibinfo {volume} {105}},\
  \bibinfo {pages} {236601} (\bibinfo {year} {2010})}\BibitemShut {NoStop}%
\bibitem [{\citenamefont {Starikov}\ \emph {et~al.}(2018)\citenamefont
  {Starikov}, \citenamefont {Liu}, \citenamefont {Yuan},\ and\ \citenamefont
  {Kelly}}]{Starikov2018}%
  \BibitemOpen
  \bibfield  {author} {\bibinfo {author} {\bibfnamefont {A.}~\bibnamefont
  {Starikov}}, \bibinfo {author} {\bibfnamefont {Y.}~\bibnamefont {Liu}},
  \bibinfo {author} {\bibfnamefont {Z.}~\bibnamefont {Yuan}},\ and\ \bibinfo
  {author} {\bibfnamefont {P.}~\bibnamefont {Kelly}},\ }\bibfield  {title}
  {\bibinfo {title} {Calculating the transport properties of magnetic materials
  from first principles including thermal and alloy disorder, noncollinearity,
  and spin-orbit coupling},\ }\href
  {https://doi.org/10.1103/PhysRevB.97.214415} {\bibfield  {journal} {\bibinfo
  {journal} {Phys. Rev. B}\ }\textbf {\bibinfo {volume} {97}},\ \bibinfo
  {pages} {214415} (\bibinfo {year} {2018})}\BibitemShut {NoStop}%
\bibitem [{\citenamefont {Garate}\ and\ \citenamefont
  {MacDonald}(2009{\natexlab{a}})}]{Garate2009a}%
  \BibitemOpen
  \bibfield  {author} {\bibinfo {author} {\bibfnamefont {I.}~\bibnamefont
  {Garate}}\ and\ \bibinfo {author} {\bibfnamefont {A.}~\bibnamefont
  {MacDonald}},\ }\bibfield  {title} {\bibinfo {title} {Gilbert damping in
  conducting ferromagnets. i. {Kohn}-{Sham} theory and atomic-scale
  inhomogeneity},\ }\href {https://doi.org/10.1103/PhysRevB.79.064403}
  {\bibfield  {journal} {\bibinfo  {journal} {Phys. Rev. B}\ }\textbf {\bibinfo
  {volume} {79}},\ \bibinfo {pages} {064403} (\bibinfo {year}
  {2009}{\natexlab{a}})}\BibitemShut {NoStop}%
\bibitem [{\citenamefont {Garate}\ and\ \citenamefont
  {MacDonald}(2009{\natexlab{b}})}]{Garate2009}%
  \BibitemOpen
  \bibfield  {author} {\bibinfo {author} {\bibfnamefont {I.}~\bibnamefont
  {Garate}}\ and\ \bibinfo {author} {\bibfnamefont {A.}~\bibnamefont
  {MacDonald}},\ }\bibfield  {title} {\bibinfo {title} {Gilbert damping in
  conducting ferromagnets. ii. model tests of the torque-correlation formula},\
  }\href {https://doi.org/10.1103/PhysRevB.79.064404} {\bibfield  {journal}
  {\bibinfo  {journal} {Phys. Rev. B}\ }\textbf {\bibinfo {volume} {79}},\
  \bibinfo {pages} {064404} (\bibinfo {year} {2009}{\natexlab{b}})}\BibitemShut
  {NoStop}%
\bibitem [{\citenamefont {Ado}\ \emph {et~al.}(2020)\citenamefont {Ado},
  \citenamefont {Ostrovsky},\ and\ \citenamefont {Titov}}]{Ado2020}%
  \BibitemOpen
  \bibfield  {author} {\bibinfo {author} {\bibfnamefont {I.~A.}\ \bibnamefont
  {Ado}}, \bibinfo {author} {\bibfnamefont {P.~M.}\ \bibnamefont {Ostrovsky}},\
  and\ \bibinfo {author} {\bibfnamefont {M.}~\bibnamefont {Titov}},\ }\bibfield
   {title} {\bibinfo {title} {Anisotropy of spin-transfer torques and {Gilbert}
  damping induced by {Rashba} coupling},\ }\href
  {https://doi.org/10.1103/PhysRevB.101.085405} {\bibfield  {journal} {\bibinfo
   {journal} {Phys. Rev. B}\ }\textbf {\bibinfo {volume} {101}},\ \bibinfo
  {pages} {085405} (\bibinfo {year} {2020})}\BibitemShut {NoStop}%
\bibitem [{\citenamefont {Gilmore}\ \emph {et~al.}(2007)\citenamefont
  {Gilmore}, \citenamefont {Idzerda},\ and\ \citenamefont
  {Stiles}}]{Gilmore2007}%
  \BibitemOpen
  \bibfield  {author} {\bibinfo {author} {\bibfnamefont {K.}~\bibnamefont
  {Gilmore}}, \bibinfo {author} {\bibfnamefont {Y.}~\bibnamefont {Idzerda}},\
  and\ \bibinfo {author} {\bibfnamefont {M.}~\bibnamefont {Stiles}},\
  }\bibfield  {title} {\bibinfo {title} {Identification of the dominant
  precession-damping mechanism in {Fe}, {Co}, and {Ni} by first-principles
  calculations},\ }\href {https://doi.org/10.1103/PhysRevLett.99.027204}
  {\bibfield  {journal} {\bibinfo  {journal} {Phys. Rev. Lett.}\ }\textbf
  {\bibinfo {volume} {99}},\ \bibinfo {pages} {027204} (\bibinfo {year}
  {2007})}\BibitemShut {NoStop}%
\bibitem [{\citenamefont {Ebert}\ \emph {et~al.}(2011)\citenamefont {Ebert},
  \citenamefont {Mankovsky}, \citenamefont {K\"odderitzsch},\ and\
  \citenamefont {Kelly}}]{Ebert2011}%
  \BibitemOpen
  \bibfield  {author} {\bibinfo {author} {\bibfnamefont {H.}~\bibnamefont
  {Ebert}}, \bibinfo {author} {\bibfnamefont {S.}~\bibnamefont {Mankovsky}},
  \bibinfo {author} {\bibfnamefont {D.}~\bibnamefont {K\"odderitzsch}},\ and\
  \bibinfo {author} {\bibfnamefont {P.~J.}\ \bibnamefont {Kelly}},\ }\bibfield
  {title} {\bibinfo {title} {Ab initio calculation of the {Gilbert} damping
  parameter via the linear response formalism},\ }\href
  {https://doi.org/10.1103/PhysRevLett.107.066603} {\bibfield  {journal}
  {\bibinfo  {journal} {Phys. Rev. Lett.}\ }\textbf {\bibinfo {volume} {107}},\
  \bibinfo {pages} {066603} (\bibinfo {year} {2011})}\BibitemShut {NoStop}%
\bibitem [{\citenamefont {Mankovsky}\ \emph {et~al.}(2013)\citenamefont
  {Mankovsky}, \citenamefont {K\"odderitzsch}, \citenamefont {Woltersdorf},\
  and\ \citenamefont {Ebert}}]{Mankovsky2013}%
  \BibitemOpen
  \bibfield  {author} {\bibinfo {author} {\bibfnamefont {S.}~\bibnamefont
  {Mankovsky}}, \bibinfo {author} {\bibfnamefont {D.}~\bibnamefont
  {K\"odderitzsch}}, \bibinfo {author} {\bibfnamefont {G.}~\bibnamefont
  {Woltersdorf}},\ and\ \bibinfo {author} {\bibfnamefont {H.}~\bibnamefont
  {Ebert}},\ }\bibfield  {title} {\bibinfo {title} {First-principles
  calculation of the {Gilbert} damping parameter via the linear response
  formalism with application to magnetic transition metals and alloys},\ }\href
  {https://doi.org/10.1103/PhysRevB.87.014430} {\bibfield  {journal} {\bibinfo
  {journal} {Phys. Rev. B}\ }\textbf {\bibinfo {volume} {87}},\ \bibinfo
  {pages} {014430} (\bibinfo {year} {2013})}\BibitemShut {NoStop}%
\bibitem [{\citenamefont {Hou}\ and\ \citenamefont {Wu}(2019)}]{Hou2019}%
  \BibitemOpen
  \bibfield  {author} {\bibinfo {author} {\bibfnamefont {Y.}~\bibnamefont
  {Hou}}\ and\ \bibinfo {author} {\bibfnamefont {R.}~\bibnamefont {Wu}},\
  }\bibfield  {title} {\bibinfo {title} {Strongly enhanced {Gilbert} damping in
  3d transition-metal ferromagnet monolayers in contact with the topological
  insulator {Bi$_2$Se$_3$}},\ }\href
  {https://doi.org/10.1103/PhysRevApplied.11.054032} {\bibfield  {journal}
  {\bibinfo  {journal} {Phys. Rev. Appl.}\ }\textbf {\bibinfo {volume} {11}},\
  \bibinfo {pages} {054032} (\bibinfo {year} {2019})}\BibitemShut {NoStop}%
\bibitem [{\citenamefont {Guimar{\~{a}}es}\ \emph {et~al.}(2019)\citenamefont
  {Guimar{\~{a}}es}, \citenamefont {Suckert}, \citenamefont {Chico},
  \citenamefont {Bouaziz}, \citenamefont {dos Santos~Dias},\ and\ \citenamefont
  {Lounis}}]{Guimaraes2019}%
  \BibitemOpen
  \bibfield  {author} {\bibinfo {author} {\bibfnamefont {F.~S.~M.}\
  \bibnamefont {Guimar{\~{a}}es}}, \bibinfo {author} {\bibfnamefont {J.~R.}\
  \bibnamefont {Suckert}}, \bibinfo {author} {\bibfnamefont {J.}~\bibnamefont
  {Chico}}, \bibinfo {author} {\bibfnamefont {J.}~\bibnamefont {Bouaziz}},
  \bibinfo {author} {\bibfnamefont {M.}~\bibnamefont {dos Santos~Dias}},\ and\
  \bibinfo {author} {\bibfnamefont {S.}~\bibnamefont {Lounis}},\ }\bibfield
  {title} {\bibinfo {title} {Comparative study of methodologies to compute the
  intrinsic {Gilbert} damping: interrelations, validity and physical
  consequences},\ }\href {https://doi.org/10.1088/1361-648x/ab1239} {\bibfield
  {journal} {\bibinfo  {journal} {J. Phys.: Condens. Matter}\ }\textbf
  {\bibinfo {volume} {31}},\ \bibinfo {pages} {255802} (\bibinfo {year}
  {2019})}\BibitemShut {NoStop}%
\bibitem [{\citenamefont {Kambersk{\'{y}}}(1976)}]{Kambersky1976}%
  \BibitemOpen
  \bibfield  {author} {\bibinfo {author} {\bibfnamefont {V.}~\bibnamefont
  {Kambersk{\'{y}}}},\ }\bibfield  {title} {\bibinfo {title} {On ferromagnetic
  resonance damping in metals},\ }\href {https://doi.org/10.1007/bf01587621}
  {\bibfield  {journal} {\bibinfo  {journal} {Czech. J. Phys.}\ }\textbf
  {\bibinfo {volume} {26}},\ \bibinfo {pages} {1366} (\bibinfo {year}
  {1976})}\BibitemShut {NoStop}%
\bibitem [{\citenamefont {Kambersk{\'{y}}}(1984)}]{Kambersky1984}%
  \BibitemOpen
  \bibfield  {author} {\bibinfo {author} {\bibfnamefont {V.}~\bibnamefont
  {Kambersk{\'{y}}}},\ }\bibfield  {title} {\bibinfo {title} {{FMR} linewidth
  and disorder in metals},\ }\href {https://doi.org/10.1007/bf01590106}
  {\bibfield  {journal} {\bibinfo  {journal} {Czech. J. Phys.}\ }\textbf
  {\bibinfo {volume} {34}},\ \bibinfo {pages} {1111} (\bibinfo {year}
  {1984})}\BibitemShut {NoStop}%
\bibitem [{\citenamefont {Kambersk\'y}(2007)}]{Kambersky2007}%
  \BibitemOpen
  \bibfield  {author} {\bibinfo {author} {\bibfnamefont {V.}~\bibnamefont
  {Kambersk\'y}},\ }\bibfield  {title} {\bibinfo {title} {Spin-orbital
  {Gilbert} damping in common magnetic metals},\ }\href
  {https://doi.org/10.1103/PhysRevB.76.134416} {\bibfield  {journal} {\bibinfo
  {journal} {Phys. Rev. B}\ }\textbf {\bibinfo {volume} {76}},\ \bibinfo
  {pages} {134416} (\bibinfo {year} {2007})}\BibitemShut {NoStop}%
\bibitem [{\citenamefont {Bar'yakhtar}(1984)}]{Baryakhtar1984}%
  \BibitemOpen
  \bibfield  {author} {\bibinfo {author} {\bibfnamefont {V.~G.}\ \bibnamefont
  {Bar'yakhtar}},\ }\bibfield  {title} {\bibinfo {title} {Phenomenological
  description of relaxation processes in magnetic materials},\ }\href@noop {}
  {\bibfield  {journal} {\bibinfo  {journal} {Sov. Phys. JETP}\ }\textbf
  {\bibinfo {volume} {60}},\ \bibinfo {pages} {863} (\bibinfo {year}
  {1984})}\BibitemShut {NoStop}%
\bibitem [{\citenamefont {Li}\ and\ \citenamefont {Bailey}(2016)}]{Li2016}%
  \BibitemOpen
  \bibfield  {author} {\bibinfo {author} {\bibfnamefont {Y.}~\bibnamefont
  {Li}}\ and\ \bibinfo {author} {\bibfnamefont {W.}~\bibnamefont {Bailey}},\
  }\bibfield  {title} {\bibinfo {title} {Wave-number-dependent {Gilbert}
  damping in metallic ferromagnets.},\ }\href
  {https://doi.org/10.1103/PhysRevLett.116.117602} {\bibfield  {journal}
  {\bibinfo  {journal} {Phys. Rev. Lett.}\ }\textbf {\bibinfo {volume} {116}},\
  \bibinfo {pages} {117602} (\bibinfo {year} {2016})}\BibitemShut {NoStop}%
\bibitem [{\citenamefont {Sayad}\ and\ \citenamefont
  {Potthoff}(2015)}]{Sayad2015}%
  \BibitemOpen
  \bibfield  {author} {\bibinfo {author} {\bibfnamefont {M.}~\bibnamefont
  {Sayad}}\ and\ \bibinfo {author} {\bibfnamefont {M.}~\bibnamefont
  {Potthoff}},\ }\bibfield  {title} {\bibinfo {title} {Spin dynamics and
  relaxation in the classical-spin {Kondo}-impurity model beyond the
  {Landau}-{Lifschitz}-{Gilbert} equation},\ }\href
  {https://doi.org/10.1088/1367-2630/17/11/113058} {\bibfield  {journal}
  {\bibinfo  {journal} {New J. Phys.}\ }\textbf {\bibinfo {volume} {17}},\
  \bibinfo {pages} {113058} (\bibinfo {year} {2015})}\BibitemShut {NoStop}%
\bibitem [{\citenamefont {Bajpai}\ and\ \citenamefont
  {Nikoli\'c}(2019)}]{Bajpai2019}%
  \BibitemOpen
  \bibfield  {author} {\bibinfo {author} {\bibfnamefont {U.}~\bibnamefont
  {Bajpai}}\ and\ \bibinfo {author} {\bibfnamefont {B.}~\bibnamefont
  {Nikoli\'c}},\ }\bibfield  {title} {\bibinfo {title} {Time-retarded damping
  and magnetic inertia in the {Landau}-{Lifshitz}-{Gilbert} equation
  self-consistently coupled to electronic time-dependent nonequilibrium {Green}
  functions},\ }\href {https://doi.org/10.1103/PhysRevB.99.134409} {\bibfield
  {journal} {\bibinfo  {journal} {Phys. Rev. B}\ }\textbf {\bibinfo {volume}
  {99}},\ \bibinfo {pages} {134409} (\bibinfo {year} {2019})}\BibitemShut
  {NoStop}%
\bibitem [{\citenamefont {Thonig}\ \emph {et~al.}(2015)\citenamefont {Thonig},
  \citenamefont {Henk},\ and\ \citenamefont {Eriksson}}]{Thonig2015}%
  \BibitemOpen
  \bibfield  {author} {\bibinfo {author} {\bibfnamefont {D.}~\bibnamefont
  {Thonig}}, \bibinfo {author} {\bibfnamefont {J.}~\bibnamefont {Henk}},\ and\
  \bibinfo {author} {\bibfnamefont {O.}~\bibnamefont {Eriksson}},\ }\bibfield
  {title} {\bibinfo {title} {Gilbert-like damping caused by time retardation in
  atomistic magnetization dynamics},\ }\href
  {https://doi.org/10.1103/PHYSREVB.92.104403} {\bibfield  {journal} {\bibinfo
  {journal} {Phys. Rev. B}\ }\textbf {\bibinfo {volume} {92}},\ \bibinfo
  {pages} {104403} (\bibinfo {year} {2015})}\BibitemShut {NoStop}%
\bibitem [{\citenamefont {Ralph}\ and\ \citenamefont
  {Stiles}(2008)}]{Ralph2008}%
  \BibitemOpen
  \bibfield  {author} {\bibinfo {author} {\bibfnamefont {D.}~\bibnamefont
  {Ralph}}\ and\ \bibinfo {author} {\bibfnamefont {M.}~\bibnamefont {Stiles}},\
  }\bibfield  {title} {\bibinfo {title} {Spin transfer torques},\ }\href
  {https://doi.org/10.1016/j.jmmm.2007.12.019} {\bibfield  {journal} {\bibinfo
  {journal} {J. Magn. Magn. Mater.}\ }\textbf {\bibinfo {volume} {320}},\
  \bibinfo {pages} {1190} (\bibinfo {year} {2008})}\BibitemShut {NoStop}%
\bibitem [{\citenamefont {Suresh}\ \emph {et~al.}(2020)\citenamefont {Suresh},
  \citenamefont {Bajpai},\ and\ \citenamefont {Nikoli{\'c}}}]{Suresh2020}%
  \BibitemOpen
  \bibfield  {author} {\bibinfo {author} {\bibfnamefont {A.}~\bibnamefont
  {Suresh}}, \bibinfo {author} {\bibfnamefont {U.}~\bibnamefont {Bajpai}},\
  and\ \bibinfo {author} {\bibfnamefont {B.~K.}\ \bibnamefont {Nikoli{\'c}}},\
  }\bibfield  {title} {\bibinfo {title} {Magnon-driven chiral charge and spin
  pumping and electron-magnon scattering from time-dependent quantum transport
  combined with classical atomistic spin dynamics},\ }\href
  {https://doi.org/10.1103/PhysRevB.101.214412} {\bibfield  {journal} {\bibinfo
   {journal} {Phys. Rev. B}\ }\textbf {\bibinfo {volume} {101}},\ \bibinfo
  {pages} {214412} (\bibinfo {year} {2020})}\BibitemShut {NoStop}%
\bibitem [{\citenamefont {Berry}\ and\ \citenamefont
  {Robbins}(1993)}]{Berry1993}%
  \BibitemOpen
  \bibfield  {author} {\bibinfo {author} {\bibfnamefont {M.~V.}\ \bibnamefont
  {Berry}}\ and\ \bibinfo {author} {\bibfnamefont {J.~M.}\ \bibnamefont
  {Robbins}},\ }\bibfield  {title} {\bibinfo {title} {Chaotic classical and
  half-classical adiabatic reactions: geometric magnetism and deterministic
  friction},\ }\href {https://doi.org/10.1098/rspa.1993.0127} {\bibfield
  {journal} {\bibinfo  {journal} {Proc. R. Soc. Lond. A: Math. Phys. Sci.}\
  }\textbf {\bibinfo {volume} {442}},\ \bibinfo {pages} {659} (\bibinfo {year}
  {1993})}\BibitemShut {NoStop}%
\bibitem [{\citenamefont {Campisi}\ \emph {et~al.}(2012)\citenamefont
  {Campisi}, \citenamefont {Denisov},\ and\ \citenamefont
  {H\"anggi}}]{Campisi2012}%
  \BibitemOpen
  \bibfield  {author} {\bibinfo {author} {\bibfnamefont {M.}~\bibnamefont
  {Campisi}}, \bibinfo {author} {\bibfnamefont {S.}~\bibnamefont {Denisov}},\
  and\ \bibinfo {author} {\bibfnamefont {P.}~\bibnamefont {H\"anggi}},\
  }\bibfield  {title} {\bibinfo {title} {Geometric magnetism in open quantum
  systems},\ }\href {https://doi.org/10.1103/PhysRevA.86.032114} {\bibfield
  {journal} {\bibinfo  {journal} {Phys. Rev. A}\ }\textbf {\bibinfo {volume}
  {86}},\ \bibinfo {pages} {032114} (\bibinfo {year} {2012})}\BibitemShut
  {NoStop}%
\bibitem [{\citenamefont {Thomas}\ \emph {et~al.}(2012)\citenamefont {Thomas},
  \citenamefont {Karzig}, \citenamefont {Kusminskiy}, \citenamefont
  {Zar\'and},\ and\ \citenamefont {von Oppen}}]{Thomas2012}%
  \BibitemOpen
  \bibfield  {author} {\bibinfo {author} {\bibfnamefont {M.}~\bibnamefont
  {Thomas}}, \bibinfo {author} {\bibfnamefont {T.}~\bibnamefont {Karzig}},
  \bibinfo {author} {\bibfnamefont {S.~V.}\ \bibnamefont {Kusminskiy}},
  \bibinfo {author} {\bibfnamefont {G.}~\bibnamefont {Zar\'and}},\ and\
  \bibinfo {author} {\bibfnamefont {F.}~\bibnamefont {von Oppen}},\ }\bibfield
  {title} {\bibinfo {title} {Scattering theory of adiabatic reaction forces due
  to out-of-equilibrium quantum environments},\ }\href
  {https://doi.org/10.1103/PhysRevB.86.195419} {\bibfield  {journal} {\bibinfo
  {journal} {Phys. Rev. B}\ }\textbf {\bibinfo {volume} {86}},\ \bibinfo
  {pages} {195419} (\bibinfo {year} {2012})}\BibitemShut {NoStop}%
\bibitem [{\citenamefont {Bajpai}\ and\ \citenamefont
  {Nikoli\'c}(2020)}]{Bajpai2020}%
  \BibitemOpen
  \bibfield  {author} {\bibinfo {author} {\bibfnamefont {U.}~\bibnamefont
  {Bajpai}}\ and\ \bibinfo {author} {\bibfnamefont {B.~K.}\ \bibnamefont
  {Nikoli\'c}},\ }\bibfield  {title} {\bibinfo {title} {Spintronics meets
  nonadiabatic molecular dynamics: Geometric spin torque and damping on
  dynamical classical magnetic texture due to an electronic open quantum
  system},\ }\href {https://doi.org/10.1103/PhysRevLett.125.187202} {\bibfield
  {journal} {\bibinfo  {journal} {Phys. Rev. Lett.}\ }\textbf {\bibinfo
  {volume} {125}},\ \bibinfo {pages} {187202} (\bibinfo {year}
  {2020})}\BibitemShut {NoStop}%
\bibitem [{\citenamefont {Kamenev}(2023)}]{Kamenev2011}%
  \BibitemOpen
  \bibfield  {author} {\bibinfo {author} {\bibfnamefont {A.}~\bibnamefont
  {Kamenev}},\ }\href {https://doi.org/10.1017/CBO9781139003667} {\emph
  {\bibinfo {title} {Field Theory of Non-Equilibrium Systems}}}\ (\bibinfo
  {publisher} {Cambridge University Press, Cambridge},\ \bibinfo {year}
  {2023})\BibitemShut {NoStop}%
\bibitem [{\citenamefont {Onoda}\ and\ \citenamefont
  {Nagaosa}(2006)}]{Onoda2006}%
  \BibitemOpen
  \bibfield  {author} {\bibinfo {author} {\bibfnamefont {M.}~\bibnamefont
  {Onoda}}\ and\ \bibinfo {author} {\bibfnamefont {N.}~\bibnamefont
  {Nagaosa}},\ }\bibfield  {title} {\bibinfo {title} {Dynamics of localized
  spins coupled to the conduction electrons with charge and spin currents},\
  }\href {https://doi.org/10.1103/PhysRevLett.96.066603} {\bibfield  {journal}
  {\bibinfo  {journal} {Phys. Rev. Lett.}\ }\textbf {\bibinfo {volume} {96}},\
  \bibinfo {pages} {066603} (\bibinfo {year} {2006})}\BibitemShut {NoStop}%
\bibitem [{\citenamefont {Rikitake}\ and\ \citenamefont
  {Imamura}(2005)}]{Rikitake2005}%
  \BibitemOpen
  \bibfield  {author} {\bibinfo {author} {\bibfnamefont {Y.}~\bibnamefont
  {Rikitake}}\ and\ \bibinfo {author} {\bibfnamefont {H.}~\bibnamefont
  {Imamura}},\ }\bibfield  {title} {\bibinfo {title} {Decoherence of localized
  spins interacting via {RKKY} interaction},\ }\href
  {https://doi.org/10.1103/PhysRevB.72.033308} {\bibfield  {journal} {\bibinfo
  {journal} {Phys. Rev. B}\ }\textbf {\bibinfo {volume} {72}},\ \bibinfo
  {pages} {033308} (\bibinfo {year} {2005})}\BibitemShut {NoStop}%
\bibitem [{\citenamefont {Fransson}(2010)}]{Fransson2010}%
  \BibitemOpen
  \bibfield  {author} {\bibinfo {author} {\bibfnamefont {J.}~\bibnamefont
  {Fransson}},\ }\bibfield  {title} {\bibinfo {title} {Dynamical exchange
  interaction between localized spins out of equilibrium},\ }\href
  {https://doi.org/10.1103/PhysRevB.82.180411} {\bibfield  {journal} {\bibinfo
  {journal} {Phys. Rev. B}\ }\textbf {\bibinfo {volume} {82}},\ \bibinfo
  {pages} {180411} (\bibinfo {year} {2010})}\BibitemShut {NoStop}%
\bibitem [{\citenamefont {N\'u{\~ n}ez}\ and\ \citenamefont
  {Duine}(2008)}]{Nunez2008}%
  \BibitemOpen
  \bibfield  {author} {\bibinfo {author} {\bibfnamefont {A.~S.}\ \bibnamefont
  {N\'u{\~ n}ez}}\ and\ \bibinfo {author} {\bibfnamefont {R.}~\bibnamefont
  {Duine}},\ }\bibfield  {title} {\bibinfo {title} {Effective temperature and
  {Gilbert} damping of a current-driven localized spin},\ }\href
  {https://doi.org/10.1103/PhysRevB.77.054401} {\bibfield  {journal} {\bibinfo
  {journal} {Phys. Rev. B}\ }\textbf {\bibinfo {volume} {77}},\ \bibinfo
  {pages} {054401} (\bibinfo {year} {2008})}\BibitemShut {NoStop}%
\bibitem [{\citenamefont {D{\'{i}}az}\ and\ \citenamefont
  {N{\'{u}}{\~{n}}ez}(2012)}]{Diaz2012}%
  \BibitemOpen
  \bibfield  {author} {\bibinfo {author} {\bibfnamefont {S.}~\bibnamefont
  {D{\'{i}}az}}\ and\ \bibinfo {author} {\bibfnamefont {{\'{A}}.~S.}\
  \bibnamefont {N{\'{u}}{\~{n}}ez}},\ }\bibfield  {title} {\bibinfo {title}
  {Current-induced exchange interactions and effective temperature in localized
  moment systems},\ }\href {https://doi.org/10.1088/0953-8984/24/11/116001}
  {\bibfield  {journal} {\bibinfo  {journal} {J. Phys.: Condens. Matter}\
  }\textbf {\bibinfo {volume} {24}},\ \bibinfo {pages} {116001} (\bibinfo
  {year} {2012})}\BibitemShut {NoStop}%
\bibitem [{\citenamefont {Leiva~M.}\ \emph {et~al.}(2023)\citenamefont
  {Leiva~M.}, \citenamefont {D\'{\i}az},\ and\ \citenamefont
  {Nunez}}]{Leiva2023}%
  \BibitemOpen
  \bibfield  {author} {\bibinfo {author} {\bibfnamefont {S.}~\bibnamefont
  {Leiva~M.}}, \bibinfo {author} {\bibfnamefont {S.~A.}\ \bibnamefont
  {D\'{\i}az}},\ and\ \bibinfo {author} {\bibfnamefont {A.~S.}\ \bibnamefont
  {Nunez}},\ }\bibfield  {title} {\bibinfo {title} {Origin of the
  magnetoelectric couplings in the spin dynamics of molecular magnets},\ }\href
  {https://doi.org/10.1103/PhysRevB.107.094401} {\bibfield  {journal} {\bibinfo
   {journal} {Phys. Rev. B}\ }\textbf {\bibinfo {volume} {107}},\ \bibinfo
  {pages} {094401} (\bibinfo {year} {2023})}\BibitemShut {NoStop}%
\bibitem [{\citenamefont {Rebei}\ \emph {et~al.}(2005)\citenamefont {Rebei},
  \citenamefont {Hitchon},\ and\ \citenamefont {Parker}}]{Rebei2005}%
  \BibitemOpen
  \bibfield  {author} {\bibinfo {author} {\bibfnamefont {A.}~\bibnamefont
  {Rebei}}, \bibinfo {author} {\bibfnamefont {W.~N.~G.}\ \bibnamefont
  {Hitchon}},\ and\ \bibinfo {author} {\bibfnamefont {G.~J.}\ \bibnamefont
  {Parker}},\ }\bibfield  {title} {\bibinfo {title}
  {$s\text{\ensuremath{-}}d$--type exchange interactions in inhomogeneous
  ferromagnets},\ }\href {https://doi.org/10.1103/PhysRevB.72.064408}
  {\bibfield  {journal} {\bibinfo  {journal} {Phys. Rev. B}\ }\textbf {\bibinfo
  {volume} {72}},\ \bibinfo {pages} {064408} (\bibinfo {year}
  {2005})}\BibitemShut {NoStop}%
\bibitem [{\citenamefont {Petrovi\'c}\ \emph {et~al.}(2018)\citenamefont
  {Petrovi\'c}, \citenamefont {Popescu}, \citenamefont {Plech\'a{\v c}},\ and\
  \citenamefont {Nikoli\'c}}]{Petrovic2018}%
  \BibitemOpen
  \bibfield  {author} {\bibinfo {author} {\bibfnamefont {M.}~\bibnamefont
  {Petrovi\'c}}, \bibinfo {author} {\bibfnamefont {B.}~\bibnamefont {Popescu}},
  \bibinfo {author} {\bibfnamefont {P.}~\bibnamefont {Plech\'a{\v c}}},\ and\
  \bibinfo {author} {\bibfnamefont {B.}~\bibnamefont {Nikoli\'c}},\ }\bibfield
  {title} {\bibinfo {title} {Spin and charge pumping by current-driven magnetic
  domain wall motion: A self-consistent multiscale
  time-dependent-quantum/time-dependent-classical approach},\ }\href
  {https://doi.org/10.1103/PhysRevApplied.10.054038} {\bibfield  {journal}
  {\bibinfo  {journal} {Phys. Rev. Appl.}\ }\textbf {\bibinfo {volume} {10}},\
  \bibinfo {pages} {054038} (\bibinfo {year} {2018})}\BibitemShut {NoStop}%
\bibitem [{\citenamefont {Petrovi{\'{c}}}\ \emph {et~al.}(2021)\citenamefont
  {Petrovi{\'{c}}}, \citenamefont {Bajpai}, \citenamefont
  {Plech{\'{a}}{\v{c}}},\ and\ \citenamefont {Nikoli{\'{c}}}}]{Petrovic2021}%
  \BibitemOpen
  \bibfield  {author} {\bibinfo {author} {\bibfnamefont {M.~D.}\ \bibnamefont
  {Petrovi{\'{c}}}}, \bibinfo {author} {\bibfnamefont {U.}~\bibnamefont
  {Bajpai}}, \bibinfo {author} {\bibfnamefont {P.}~\bibnamefont
  {Plech{\'{a}}{\v{c}}}},\ and\ \bibinfo {author} {\bibfnamefont {B.~K.}\
  \bibnamefont {Nikoli{\'{c}}}},\ }\bibfield  {title} {\bibinfo {title}
  {Annihilation of topological solitons in magnetism with spin-wave burst
  finale: Role of nonequilibrium electrons causing nonlocal damping and spin
  pumping over ultrabroadband frequency range},\ }\href
  {https://doi.org/10.1103/PhysRevB.104.L020407} {\bibfield  {journal}
  {\bibinfo  {journal} {Phys. Rev. B}\ }\textbf {\bibinfo {volume} {104}},\
  \bibinfo {pages} {l020407} (\bibinfo {year} {2021})}\BibitemShut {NoStop}%
\bibitem [{\citenamefont {Costa}\ and\ \citenamefont
  {Muniz}(2015)}]{Costa2015}%
  \BibitemOpen
  \bibfield  {author} {\bibinfo {author} {\bibfnamefont {A.~T.}\ \bibnamefont
  {Costa}}\ and\ \bibinfo {author} {\bibfnamefont {R.~B.}\ \bibnamefont
  {Muniz}},\ }\bibfield  {title} {\bibinfo {title} {Breakdown of the adiabatic
  approach for magnetization damping in metallic ferromagnets},\ }\href
  {https://doi.org/10.1103/PhysRevB.92.014419} {\bibfield  {journal} {\bibinfo
  {journal} {Phys. Rev. B}\ }\textbf {\bibinfo {volume} {92}},\ \bibinfo
  {pages} {014419} (\bibinfo {year} {2015})}\BibitemShut {NoStop}%
\bibitem [{\citenamefont {Edwards}(2016)}]{Edwards2016}%
  \BibitemOpen
  \bibfield  {author} {\bibinfo {author} {\bibfnamefont {D.~M.}\ \bibnamefont
  {Edwards}},\ }\bibfield  {title} {\bibinfo {title} {The absence of intraband
  scattering in a consistent theory of {Gilbert} damping in pure metallic
  ferromagnets},\ }\href {http://stacks.iop.org/0953-8984/28/i=8/a=086004}
  {\bibfield  {journal} {\bibinfo  {journal} {J. Phys.: Condens. Matter}\
  }\textbf {\bibinfo {volume} {28}},\ \bibinfo {pages} {086004} (\bibinfo
  {year} {2016})}\BibitemShut {NoStop}%
\bibitem [{\citenamefont {Mahfouzi}\ \emph {et~al.}(2017)\citenamefont
  {Mahfouzi}, \citenamefont {Kim},\ and\ \citenamefont
  {Kioussis}}]{Mahfouzi2017a}%
  \BibitemOpen
  \bibfield  {author} {\bibinfo {author} {\bibfnamefont {F.}~\bibnamefont
  {Mahfouzi}}, \bibinfo {author} {\bibfnamefont {J.}~\bibnamefont {Kim}},\ and\
  \bibinfo {author} {\bibfnamefont {N.}~\bibnamefont {Kioussis}},\ }\bibfield
  {title} {\bibinfo {title} {Intrinsic damping phenomena from quantum to
  classical magnets: An ab initio study of {Gilbert} damping in a {Pt/Co}
  bilayer},\ }\href {https://doi.org/10.1103/PhysRevB.96.214421} {\bibfield
  {journal} {\bibinfo  {journal} {Phys. Rev. B}\ }\textbf {\bibinfo {volume}
  {96}},\ \bibinfo {pages} {214421} (\bibinfo {year} {2017})}\BibitemShut
  {NoStop}%
\bibitem [{\citenamefont {Tserkovnyak}\ \emph {et~al.}(2005)\citenamefont
  {Tserkovnyak}, \citenamefont {Brataas}, \citenamefont {Bauer},\ and\
  \citenamefont {Halperin}}]{Tserkovnyak2005}%
  \BibitemOpen
  \bibfield  {author} {\bibinfo {author} {\bibfnamefont {Y.}~\bibnamefont
  {Tserkovnyak}}, \bibinfo {author} {\bibfnamefont {A.}~\bibnamefont
  {Brataas}}, \bibinfo {author} {\bibfnamefont {G.~E.~W.}\ \bibnamefont
  {Bauer}},\ and\ \bibinfo {author} {\bibfnamefont {B.~I.}\ \bibnamefont
  {Halperin}},\ }\bibfield  {title} {\bibinfo {title} {Nonlocal magnetization
  dynamics in ferromagnetic heterostructures},\ }\href
  {https://doi.org/10.1103/RevModPhys.77.1375} {\bibfield  {journal} {\bibinfo
  {journal} {Rev. Mod. Phys.}\ }\textbf {\bibinfo {volume} {77}},\ \bibinfo
  {pages} {1375} (\bibinfo {year} {2005})}\BibitemShut {NoStop}%
\bibitem [{\citenamefont {Brataas}\ \emph {et~al.}(2011)\citenamefont
  {Brataas}, \citenamefont {Tserkovnyak},\ and\ \citenamefont
  {Bauer}}]{Brataas2011}%
  \BibitemOpen
  \bibfield  {author} {\bibinfo {author} {\bibfnamefont {A.}~\bibnamefont
  {Brataas}}, \bibinfo {author} {\bibfnamefont {Y.}~\bibnamefont
  {Tserkovnyak}},\ and\ \bibinfo {author} {\bibfnamefont {G.~E.~W.}\
  \bibnamefont {Bauer}},\ }\bibfield  {title} {\bibinfo {title} {Magnetization
  dissipation in ferromagnets from scattering theory},\ }\href
  {https://doi.org/10.1103/PhysRevB.84.054416} {\bibfield  {journal} {\bibinfo
  {journal} {Phys. Rev. B}\ }\textbf {\bibinfo {volume} {84}},\ \bibinfo
  {pages} {054416} (\bibinfo {year} {2011})}\BibitemShut {NoStop}%
\bibitem [{\citenamefont {Nagaosa}\ \emph {et~al.}(2009)\citenamefont
  {Nagaosa}, \citenamefont {Sinova}, \citenamefont {Onoda}, \citenamefont
  {Mac{D}onald},\ and\ \citenamefont {Ong}}]{Nagaosa2009}%
  \BibitemOpen
  \bibfield  {author} {\bibinfo {author} {\bibfnamefont {N.}~\bibnamefont
  {Nagaosa}}, \bibinfo {author} {\bibfnamefont {J.}~\bibnamefont {Sinova}},
  \bibinfo {author} {\bibfnamefont {S.}~\bibnamefont {Onoda}}, \bibinfo
  {author} {\bibfnamefont {A.~H.}\ \bibnamefont {Mac{D}onald}},\ and\ \bibinfo
  {author} {\bibfnamefont {N.}~\bibnamefont {Ong}},\ }\bibfield  {title}
  {\bibinfo {title} {Anomalous {Hall} effect},\ }\href
  {https://doi.org/10.1103/RevModPhys.82.1539} {\bibfield  {journal} {\bibinfo
  {journal} {Rev. Mod. Phys.}\ }\textbf {\bibinfo {volume} {82}},\ \bibinfo
  {pages} {1539} (\bibinfo {year} {2009})}\BibitemShut {NoStop}%
\bibitem [{\citenamefont {Altland}\ and\ \citenamefont
  {Simons}(2023)}]{Altland2010}%
  \BibitemOpen
  \bibfield  {author} {\bibinfo {author} {\bibfnamefont {A.}~\bibnamefont
  {Altland}}\ and\ \bibinfo {author} {\bibfnamefont {B.}~\bibnamefont
  {Simons}},\ }\href {https://doi.org/10.1017/CBO9780511789984.012} {\emph
  {\bibinfo {title} {Condensed Matter Field Theory}}}\ (\bibinfo  {publisher}
  {Cambridge University Press, Cambridge},\ \bibinfo {year} {2023})\BibitemShut
  {NoStop}%
\bibitem [{\citenamefont {Shnirman}\ \emph {et~al.}(2015)\citenamefont
  {Shnirman}, \citenamefont {Gefen}, \citenamefont {Saha}, \citenamefont
  {Burmistrov}, \citenamefont {Kiselev},\ and\ \citenamefont
  {Altland}}]{Shnirman2015}%
  \BibitemOpen
  \bibfield  {author} {\bibinfo {author} {\bibfnamefont {A.}~\bibnamefont
  {Shnirman}}, \bibinfo {author} {\bibfnamefont {Y.}~\bibnamefont {Gefen}},
  \bibinfo {author} {\bibfnamefont {A.}~\bibnamefont {Saha}}, \bibinfo {author}
  {\bibfnamefont {I.}~\bibnamefont {Burmistrov}}, \bibinfo {author}
  {\bibfnamefont {M.}~\bibnamefont {Kiselev}},\ and\ \bibinfo {author}
  {\bibfnamefont {A.}~\bibnamefont {Altland}},\ }\bibfield  {title} {\bibinfo
  {title} {Geometric quantum noise of spin},\ }\href
  {https://doi.org/10.1103/PhysRevLett.114.176806} {\bibfield  {journal}
  {\bibinfo  {journal} {Phys. Rev. Lett.}\ }\textbf {\bibinfo {volume} {114}},\
  \bibinfo {pages} {176806} (\bibinfo {year} {2015})}\BibitemShut {NoStop}%
\bibitem [{\citenamefont {Verstraten}\ \emph {et~al.}(2023)\citenamefont
  {Verstraten}, \citenamefont {Ludwig}, \citenamefont {Duine},\ and\
  \citenamefont {Morais~Smith}}]{Verstraten2023}%
  \BibitemOpen
  \bibfield  {author} {\bibinfo {author} {\bibfnamefont {R.~C.}\ \bibnamefont
  {Verstraten}}, \bibinfo {author} {\bibfnamefont {T.}~\bibnamefont {Ludwig}},
  \bibinfo {author} {\bibfnamefont {R.~A.}\ \bibnamefont {Duine}},\ and\
  \bibinfo {author} {\bibfnamefont {C.}~\bibnamefont {Morais~Smith}},\
  }\bibfield  {title} {\bibinfo {title} {Fractional {Landau-Lifshitz-Gilbert}
  equation},\ }\href {https://doi.org/10.1103/PhysRevResearch.5.033128}
  {\bibfield  {journal} {\bibinfo  {journal} {Phys. Rev. Res.}\ }\textbf
  {\bibinfo {volume} {5}},\ \bibinfo {pages} {033128} (\bibinfo {year}
  {2023})}\BibitemShut {NoStop}%
\bibitem [{\citenamefont {Hurst}\ \emph {et~al.}(2020)\citenamefont {Hurst},
  \citenamefont {Galitski},\ and\ \citenamefont {Heikkil\"a}}]{Hurst2020}%
  \BibitemOpen
  \bibfield  {author} {\bibinfo {author} {\bibfnamefont {H.~M.}\ \bibnamefont
  {Hurst}}, \bibinfo {author} {\bibfnamefont {V.}~\bibnamefont {Galitski}},\
  and\ \bibinfo {author} {\bibfnamefont {T.~T.}\ \bibnamefont {Heikkil\"a}},\
  }\bibfield  {title} {\bibinfo {title} {Electron-induced massive dynamics of
  magnetic domain walls},\ }\href {https://doi.org/10.1103/PhysRevB.101.054407}
  {\bibfield  {journal} {\bibinfo  {journal} {Phys. Rev. B}\ }\textbf {\bibinfo
  {volume} {101}},\ \bibinfo {pages} {054407} (\bibinfo {year}
  {2020})}\BibitemShut {NoStop}%
\bibitem [{\citenamefont {Manchon}\ \emph {et~al.}(2015)\citenamefont
  {Manchon}, \citenamefont {Koo}, \citenamefont {Nitta}, \citenamefont
  {Frolov},\ and\ \citenamefont {Duine}}]{Manchon2015}%
  \BibitemOpen
  \bibfield  {author} {\bibinfo {author} {\bibfnamefont {A.}~\bibnamefont
  {Manchon}}, \bibinfo {author} {\bibfnamefont {H.~C.}\ \bibnamefont {Koo}},
  \bibinfo {author} {\bibfnamefont {J.}~\bibnamefont {Nitta}}, \bibinfo
  {author} {\bibfnamefont {S.~M.}\ \bibnamefont {Frolov}},\ and\ \bibinfo
  {author} {\bibfnamefont {R.~A.}\ \bibnamefont {Duine}},\ }\bibfield  {title}
  {\bibinfo {title} {New perspectives for {Rashba} spin{\textendash}orbit
  coupling},\ }\href {https://doi.org/10.1038/nmat4360} {\bibfield  {journal}
  {\bibinfo  {journal} {Nat. Mater.}\ }\textbf {\bibinfo {volume} {14}},\
  \bibinfo {pages} {871} (\bibinfo {year} {2015})}\BibitemShut {NoStop}%
\bibitem [{\citenamefont {Baker}\ \emph {et~al.}(2016)\citenamefont {Baker},
  \citenamefont {Figueroa}, \citenamefont {Love}, \citenamefont {Cavill},
  \citenamefont {Hesjedal},\ and\ \citenamefont {van~der Laan}}]{Baker2016}%
  \BibitemOpen
  \bibfield  {author} {\bibinfo {author} {\bibfnamefont {A.}~\bibnamefont
  {Baker}}, \bibinfo {author} {\bibfnamefont {A.~I.}\ \bibnamefont {Figueroa}},
  \bibinfo {author} {\bibfnamefont {C.}~\bibnamefont {Love}}, \bibinfo {author}
  {\bibfnamefont {S.}~\bibnamefont {Cavill}}, \bibinfo {author} {\bibfnamefont
  {T.}~\bibnamefont {Hesjedal}},\ and\ \bibinfo {author} {\bibfnamefont
  {G.}~\bibnamefont {van~der Laan}},\ }\bibfield  {title} {\bibinfo {title}
  {Anisotropic absorption of pure spin currents},\ }\href
  {https://doi.org/10.1103/PhysRevLett.116.047201} {\bibfield  {journal}
  {\bibinfo  {journal} {Phys. Rev. Lett.}\ }\textbf {\bibinfo {volume} {116}},\
  \bibinfo {pages} {047201} (\bibinfo {year} {2016})}\BibitemShut {NoStop}%
\bibitem [{\citenamefont {F\"ahnle}\ \emph {et~al.}(2008)\citenamefont
  {F\"ahnle}, \citenamefont {Steiauf},\ and\ \citenamefont
  {Seib}}]{Faehnle2008}%
  \BibitemOpen
  \bibfield  {author} {\bibinfo {author} {\bibfnamefont {M.}~\bibnamefont
  {F\"ahnle}}, \bibinfo {author} {\bibfnamefont {D.}~\bibnamefont {Steiauf}},\
  and\ \bibinfo {author} {\bibfnamefont {J.}~\bibnamefont {Seib}},\ }\bibfield
  {title} {\bibinfo {title} {The {Gilbert} equation revisited: anisotropic and
  nonlocal damping of magnetization dynamics},\ }\href
  {https://doi.org/10.1088/0022-3727/41/16/164014} {\bibfield  {journal}
  {\bibinfo  {journal} {J. Phys. D: Appl. Phys.}\ }\textbf {\bibinfo {volume}
  {41}},\ \bibinfo {pages} {164014} (\bibinfo {year} {2008})}\BibitemShut
  {NoStop}%
\bibitem [{\citenamefont {Montoya}\ \emph {et~al.}(2014)\citenamefont
  {Montoya}, \citenamefont {Heinrich},\ and\ \citenamefont
  {Girt}}]{Montoya2014}%
  \BibitemOpen
  \bibfield  {author} {\bibinfo {author} {\bibfnamefont {E.}~\bibnamefont
  {Montoya}}, \bibinfo {author} {\bibfnamefont {B.}~\bibnamefont {Heinrich}},\
  and\ \bibinfo {author} {\bibfnamefont {E.}~\bibnamefont {Girt}},\ }\bibfield
  {title} {\bibinfo {title} {Quantum well state induced oscillation of pure
  spin currents in {Fe/Au/Pd}(001) systems},\ }\href
  {https://doi.org/10.1103/PHYSREVLETT.113.136601} {\bibfield  {journal}
  {\bibinfo  {journal} {Phys. Rev. Lett.}\ }\textbf {\bibinfo {volume} {113}},\
  \bibinfo {pages} {136601} (\bibinfo {year} {2014})}\BibitemShut {NoStop}%
\bibitem [{\citenamefont {Ryndyk}(2016)}]{Ryndyk2016}%
  \BibitemOpen
  \bibfield  {author} {\bibinfo {author} {\bibfnamefont {D.}~\bibnamefont
  {Ryndyk}},\ }\href
  {https://www.semanticscholar.org/paper/d10c821cc8a03c7747ec1bac4a00782537f4e5d8}
  {\emph {\bibinfo {title} {Theory of Quantum Transport at Nanoscale}}}\
  (\bibinfo  {publisher} {Springer, Cham},\ \bibinfo {year} {2016})\BibitemShut
  {NoStop}%
\bibitem [{\citenamefont {Barnes}\ and\ \citenamefont
  {Maekawa}(2007)}]{Barnes2007}%
  \BibitemOpen
  \bibfield  {author} {\bibinfo {author} {\bibfnamefont {S.~E.}\ \bibnamefont
  {Barnes}}\ and\ \bibinfo {author} {\bibfnamefont {S.}~\bibnamefont
  {Maekawa}},\ }\bibfield  {title} {\bibinfo {title} {Generalization of
  {Faraday's} law to include nonconservative spin forces},\ }\href
  {https://doi.org/10.1103/PhysRevLett.98.246601} {\bibfield  {journal}
  {\bibinfo  {journal} {Phys. Rev. Lett.}\ }\textbf {\bibinfo {volume} {98}},\
  \bibinfo {pages} {246601} (\bibinfo {year} {2007})}\BibitemShut {NoStop}%
\bibitem [{\citenamefont {Duine}(2009)}]{Duine2009}%
  \BibitemOpen
  \bibfield  {author} {\bibinfo {author} {\bibfnamefont {R.~A.}\ \bibnamefont
  {Duine}},\ }\bibfield  {title} {\bibinfo {title} {Effects of nonadiabaticity
  on the voltage generated by a moving domain wall},\ }\href
  {https://doi.org/10.1103/PhysRevB.79.014407} {\bibfield  {journal} {\bibinfo
  {journal} {Phys. Rev. B}\ }\textbf {\bibinfo {volume} {79}},\ \bibinfo
  {pages} {014407} (\bibinfo {year} {2009})}\BibitemShut {NoStop}%
\bibitem [{\citenamefont {Tatara}(2019)}]{Tatara2019}%
  \BibitemOpen
  \bibfield  {author} {\bibinfo {author} {\bibfnamefont {G.}~\bibnamefont
  {Tatara}},\ }\bibfield  {title} {\bibinfo {title} {Effective gauge field
  theory of spintronics},\ }\href {https://doi.org/10.1016/j.physe.2018.05.011}
  {\bibfield  {journal} {\bibinfo  {journal} {Phys. E: Low-Dimens. Syst.
  Nanostructures}\ }\textbf {\bibinfo {volume} {106}},\ \bibinfo {pages} {208}
  (\bibinfo {year} {2019})}\BibitemShut {NoStop}%
\bibitem [{\citenamefont {Cooper}\ and\ \citenamefont
  {Uehling}(1967)}]{Cooper1967}%
  \BibitemOpen
  \bibfield  {author} {\bibinfo {author} {\bibfnamefont {R.~L.}\ \bibnamefont
  {Cooper}}\ and\ \bibinfo {author} {\bibfnamefont {E.~A.}\ \bibnamefont
  {Uehling}},\ }\bibfield  {title} {\bibinfo {title} {Ferromagnetic resonance
  and spin diffusion in supermalloy},\ }\href
  {https://doi.org/10.1103/PhysRev.164.662} {\bibfield  {journal} {\bibinfo
  {journal} {Phys. Rev.}\ }\textbf {\bibinfo {volume} {164}},\ \bibinfo {pages}
  {662} (\bibinfo {year} {1967})}\BibitemShut {NoStop}%
\end{thebibliography}%

\end{document}